\documentclass[preprint,showpacs,preprintnumbers,amsmath,amssymb,showkeys,nofootinbib]{revtex4}

\usepackage[cp1251]{inputenc}

\usepackage{graphicx}
\usepackage{dcolumn}
\usepackage{bm}

\usepackage{float}
\usepackage{stmaryrd} 

\usepackage{version}

\usepackage[unicode,bookmarks,bookmarksopen,bookmarksopenlevel=0,colorlinks,%
pageanchor=true,breaklinks=true,linkcolor=red,citecolor=blue,urlcolor=red,
pdfencoding = auto, bookmarksdepth = 4]{hyperref}  

\newcommand{\const}{\textrm{const}}

\begin{document}


\title{Hodograph Method and Numerical Solution of the \\ Two Hyperbolic Quasilinear  Equations System. \\ Part II. Zonal Electrophoresis Equations}

\author{E.\,V.~Shiryaeva}%
 \email{shir@math.sfedu.ru}
\affiliation{%
Institute of Mathematics, Mechanics and Computer Science, Southern Federal University, Rostov-on-Don, Russia
}%

\author{M.\,Yu.~Zhukov}
\email{zhuk@math.sdedu.ru}
\affiliation{%
Institute of Mathematics, Mechanics and Computer Science, Southern Federal University, Rostov-on-Don, Russia
}%

\date{\today}

\begin{abstract}

The paper presents the solutions for the  zonal electrophoresis equations are obtained by analytical and numerical methods. The method proposed by the authors is used. This method allows to reduce the Cauchy problem for two hyperbolic quasilinear  PDE's to the Cauchy problem for ODE's. In some respect, this method is analogous to the method of characteristics for two hyperbolic equations. The method is effectively applicable in all cases when the explicit expression for the  Riemann--Green function of some linear second order PDE, resulting from the use of the hodograph method for the original equations, is known. One of the method advantages is the possibility of constructing a multi-valued solutions. Compared with the previous authors paper, in which, in particular, the  shallow water equations are studied, here we investigate the case when the  Riemann--Green function can be represent as
the sum of the terms  each of them is a product of two multipliers depended on different variables.
The numerical results for zonal electrophoresis equations are presented. For computing the different initial data (periodic, wave packet, the Gaussian distribution) are used.

\end{abstract}

\pacs{02.30.Jr, 02.30.Hq, 82.45.-h, 87.15.Tt, 87.50.ch, 02.60.-x}





\keywords{hodograph method, hyperbolic quasilinear equations, zonal electrophoresis equations}
\maketitle

\section{Introduction}\label{zhshel:sec:introd}

In previous paper \cite{Zhuk_Shir_ArXiv_2014_1}, the efficient numerical method, allowing to get solutions, including multivalued\footnote{In \cite{Zhuk_Shir_ArXiv_2014_1} the solutions of the shallow water equations describing the braking waves are presented.}, are proposed in the case of the Cauchy problem for two hyperbolic quasilinear  PDE's. This method is based on the results of the paper \cite{SenashovYakhno} in which the hodograph method based on
conservation laws for the two quasilinear hyperbolic PDE's is presented.
For the determination of the densities and fluxes of some conservation laws the linear hyperbolic  second order PDE is used.
As shown in \cite{SenashovYakhno},  the solution of the original equations can easily be written in implicit analytical form if
there is an analytical expression for the  Riemann--Green function of mentioned linear hyperbolic equation.

The paper \cite{Zhuk_Shir_ArXiv_2014_1} shows that one can not only write the solution in  implicit analytical form, but also construct the efficient numerical method of the Cauchy problem integration. Using minor modifications of the results of \cite{SenashovYakhno} it is able to reduce the Cauchy problem for two quasilinear PDE's to the Cauchy problem for ODE's.
From the authors point of view, the solution of the Cauchy problem for ODE's, in particular, numerical solution,  is much easier than the solution of nonlinear transcendental equations that must be solved when there is an implicit solution of the original problem.

A key role for the proposed method plays the possibility of constructing an explicit expression  for the Riemann--Green function of the corresponding linear equation. This, of course, limits the application of the method. In fact, the number of equations to which the method is applicable is large enough. These include the shallow water equations (see, for example, \cite{RozhdestvenskiiYanenko,Whithem}), the gas dynamics equations for a polytropic gas \cite{RozhdestvenskiiYanenko,Whithem}, the soliton gas equations \cite{Whithem,GenaEl} (or Born--Infeld equation), the chromatography equations  for classical isotherms \cite{RozhdestvenskiiYanenko,FerapontovTsarev_MatModel,Kuznetsov}, the isotachophoresis and the zonal electrophoresis equations \cite{BabskiiZhukovYudovichRussian,ZhukovMassTransport,ZhukovNonSteadyITP,ElaevaMM,Elaeva_ZhVM}. A~large number of equations, for which the  explicit expression for the Riemann--Green functions are known, are presented, in particular, in \cite{SenashovYakhno}.
Classification of equations that allow explicit expressions for the  Riemann--Green functions, is contained in \cite{Copson,Courant,Ibragimov} (see also \cite{Chirkunov,Chirkunov_2}).

This paper presents analytical and numerical solution of the Cauchy problem for the zonal electrophoresis equations \cite{BabskiiZhukovYudovichRussian,ZhukovMassTransport,ZhukovNonSteadyITP,ElaevaMM,Elaeva_ZhVM}. The choice of these problem, in particular,  due to the fact that  there is the Riemann--Green function which can be represented as a finite sum of two multipliers each of them depends on the different variable
\begin{equation}\label{zhshArXiv_Part_2:eq:1.01}
\Phi(R^1,R^2|r^1,r^2)=\sum\limits_{k=1}^{n}\mathcal{P}_k(R^1,R^2)\mathcal{Q}_k(r^1,r^2).
\end{equation}
It is shown below, that this type of function allows to significantly simplify the construction of solutions.

Pay attention to the fact that in some sense, the proposed method is `exact'. Its realization  does not require any approximation of the original hyperbolic PDE's, which uses of the finite-difference methods, finite element method, finite volume method, the Riemann solver, etc.
Also there is no need to introduce an artificial viscosity\footnote{The effect of the grid viscosity does not occur due to the absence of approximation}. In other words, the original problem is solved without any approximations or modifications. The accuracy of the solution is determined by only the accuracy of the ODE's numerical solution method.

The paper is organized as follows. In Secs.~\ref{zhshArXiv:sec:2} we repeat the slightly modified (and simplified) results of the paper \cite{SenashovYakhno}.
In Sec.~\ref{zhshArXiv:sec:2} we formulate the problem for the zonal electrophoresis equations. Also we construct the Cauchy ODE's problem which allows to obtain the solution of the original problem on the isochrones. In this section we present the results of calculating for the different initial data.

\setcounter{equation}{0}

\section{Reduction of the Cauchy problem for two hyperbolic quasilinear  PDE's to the Cauchy problem for ODE's}\label{zhshArXiv:sec:2}

Referring for details to \cite{Zhuk_Shir_ArXiv_2014_1,SenashovYakhno}, here we give only a brief description of the method which allows to reduce
the Cauchy problem for two hyperbolic quasilinear PDE's to Cauchy problem for ODE's.


\subsection{The Riemann invariants}\label{zhshArXiv:sec:2.A}

Let for a system of two hyperbolic PDE's, written in the Riemann invariants $R^1(x,t)$, $R^2(x,t)$,  we have the Cauchy problem at $t=t_0$
\begin{equation}\label{zhshArXiv:eq:2.01}
R^1_t+ \lambda^1(R^1,R^2)R^1_x=0, \quad R^2_t+ \lambda^2(R^1,R^2)R^2_x=0,
\end{equation}
\begin{equation}\label{zhshArXiv:eq:2.02}
R^1(\tau,t_0)=R^1_0(x), \quad R^2(x,t_0)=R^2_0(x),
\end{equation}
where $R^1_0(x)$, $R^2_0(x)$ are the functions determined on some interval of the axis $x$  (possibly infinite), $\lambda^1(R^1,R^2)$, $\lambda^2(R^1,R^2)$ are the given characteristic directions.

\subsection{Hodograph method}\label{zhshArXiv:sec:2.B}

Using the hodograph method for some conservation law
$\varphi_t+\psi_x=0$, where $\varphi(R^1,R^2)$ is the density, $\psi(R^1,R^2)$ is the flux, we write the equation \cite{SenashovYakhno}
\begin{equation}\label{zhshArXiv:eq:2.03}
\Phi_{R^1 R^2} + A(R^1,R^2)\Phi_{R^1} + B(R^1,R^2)\Phi_{R^2}=0,
\end{equation}
\begin{equation}\label{zhshArXiv:eq:2.04}
A(R^1,R^2)=\frac{\lambda^1_{R^2}}{\lambda^1-\lambda^2}, \quad
B(R^1,R^2)=-\frac{\lambda^2_{R^1}}{\lambda^1-\lambda^2}.
\end{equation}

\subsection{The Riemann--Green function}\label{zhshArXiv:sec:2.C}

Let the function $\Phi(R^1,R^2|r^1,r^2)$ be the Riemann--Green function  for equation (\ref{zhshArXiv:eq:2.03}).
The function $\Phi(R^1,R^2|r^1,r^2)$ of variables $R^1$, $R^2$ satisfies the given equation, and
the function $\Phi(R^1,R^2|r^1,r^2)$ of variables $r^1$, $r^2$ is the solution of the conjugate problem
\begin{equation}\label{zhshArXiv:eq:2.05}
\Phi_{r^1 r^2} - (A(r^1,r^2)\Phi)_{r^1} - (B(r^1,r^2)\Phi)_{r^2}=0,
\end{equation}
\begin{equation}\label{zhshArXiv:eq:2.06}
(\Phi_{r^2} - A\Phi)\bigr|_{r^1=R^1}=0, \quad (\Phi_{r^1} - B\Phi)\bigr|_{r^2=R^2}=0,
\end{equation}
\begin{equation}\label{zhshArXiv:eq:2.07}
\Phi\bigr|_{r^1=R^1,r^2=R^2}=1.
\end{equation}
The construction methods of the Riemann--Green function  are described, for example, in \cite{Copson,Chirkunov,Chirkunov_2,Courant,Ibragimov,SenashovYakhno}.

\subsection{Implicit solution of the problem}\label{zhshArXiv:sec:2.D}

It is convenient, to write the density of a conservation law, i.e. the function $\varphi(R^1,R^2)$,
 in the form $\varphi(R^1,R^2|r^1,r^2)$
\begin{equation}\label{zhshArXiv:eq:2.08}
  \varphi(R^1,R^2|r^1,r^2) = M(r^1,r^2)\Phi(R^1,R^2|r^1,r^2), \quad   M(r^1,r^2)=\frac{2}{\lambda^2(r^1,r^2)-\lambda^1(r^1,r^2)}.
\end{equation}

The solution of (\ref{zhshArXiv:eq:2.01}), (\ref{zhshArXiv:eq:2.02}) can be represented in implicit form as \cite{SenashovYakhno}
\begin{equation}\label{zhshArXiv:eq:2.09}
R^1(x,t)=r^1(b)=R^1_0(b), \quad R^2(x,t)=r^2(a)=R^2_0(a),
\end{equation}
where $a$, $b$ are the new variables (Lagrangian variables).

The connection between the new variables $a$, $b$ and old variables $x$, $t$ has the form
\begin{equation}\label{zhshArXiv:eq:2.10}
t=t(a,b), \quad x=x(a,b).
\end{equation}

Function $t=t(a,b)$ is calculated using the density of the conservation law $\varphi(R^1,R^2|r^1,r^2)$ and the initial data $R^1_0(x)$, $R^2_0(x)$ \cite{SenashovYakhno,Zhuk_Shir_ArXiv_2014_1}
\begin{equation}\label{zhshArXiv:eq:2.11}
t(a,b)=t_0+\frac12\int\limits_{a}^{b}\varphi(R^1_0(\tau),R^2_0(\tau)|r^1(b),r^2(a))\,d\tau.
\end{equation}

Function $x=x(a,b)$ is calculated by analogy \cite{SenashovYakhno}. Note, that this function is not required for further. We assume that this function is the given function.

If the equations (\ref{zhshArXiv:eq:2.10}) are solvable explicitly
\begin{equation}\label{zhshArXiv:eq:2.12}
  a=a(x,t), \quad b=b(x,t),
\end{equation}
then we have explicit solution
\begin{equation}\label{zhshArXiv:eq:2.13}
R^1(x,t)=R^1_0(b(x,t)), \quad R^2(x,t)=R^2_0(a(x,t)).
\end{equation}

In principle, one can assume that the original problem is solved. There is a system of nonlinear transcendental equations (\ref{zhshArXiv:eq:2.10}) for variables $a$, $b$, where the functions $t(a,b)$, $x(a,b)$ are completely determined. Solving this system for each fixed point $(x,t)$ we obtain the solution in the form (\ref{zhshArXiv:eq:2.12}). If we have good numerical algorithms for solving systems of transcendental equations and good initial approximations, then  the solution of the Cauchy problem for ODE's is not required.

\subsection{Solution on isochrones}\label{zhshArXiv:sec:2.E}

To construct the solution in the form (\ref{zhshArXiv:eq:2.09}) we proposed \cite{Zhuk_Shir_ArXiv_2014_1} to solve the Cauchy problem for ODE's.
We fix some value $t=t_*$, specifying the level line (isochrone) of function $t(a,b)$
\begin{equation}\label{zhshArXiv:eq:2.14}
t_*=t(a, b).
\end{equation}
We assume that the isochrone is determined on the plane $(a,b)$  by the parametrical equations
\begin{equation}\label{zhshArXiv:eq:2.15}
a=a(\tau), \quad b=b(\tau),
\end{equation}
where $\tau$ is the parameter.

We choose the values $a_*$, $b_*$ which determine some point on isochrone $t=t_*$
\begin{equation}\label{zhshArXiv:eq:2.16}
t_*=t(a_*, b_*).
\end{equation}
In practice, the values of $a_*$, $b_*$ one can choose using the  line levels of function $t(a,b)$ for some ranges of parameters $a$, $b$.

The coordinate $x$ on isochrone, obviously, is determined by the expression
\begin{equation}\label{zhshArXiv:eq:2.17}
x=x(a(\tau), b(\tau))\equiv X(\tau).
\end{equation}

To determine the functions $a(\tau)$, $b(\tau)$, $X(\tau)$ we have the  Cauchy problem \cite{Zhuk_Shir_ArXiv_2014_1}
\begin{equation}\label{zhshArXiv:eq:2.18}
\frac{da}{d\tau}=-t_b(a,b), \quad
\frac{db}{d\tau}=t_a(a,b),
\end{equation}
\begin{equation}\label{zhshArXiv:eq:2.19}
\frac{dX}{d\tau}=(\lambda^2(r^1(b),r^2(a))-\lambda^1(r^1(b),r^2(a)))t_a(a,b) t_b(a,b),
\end{equation}
\begin{equation}\label{zhshArXiv:eq:2.20}
a\bigr|_{\tau=0}=a_*, \quad b\bigr|_{\tau=0}=b_*, \quad
X\bigr|_{\tau=0}=X_*.
\end{equation}
Here the values $a_*$, $b_*$ are given. To determine  $X_*$ we need to solve the problem
\begin{equation}\label{zhshArXiv:eq:2.21}
\frac{dY(b)}{db}=x_b(a_*,b)=\lambda^2(r^1(b),r^2(a_*))t_b(a_*,b), \quad Y(a_*)=a_*.
\end{equation}
Integrating from $a_*$ to $b_*$ we get
\begin{equation}\label{zhshArXiv:eq:2.22}
X_*=Y(b_*).
\end{equation}
Note, that $X_*=x(a_*,b_*)$ is the $x$ coordinate corresponding to $\tau=0$.

Integrating the Cauchy problem (\ref{zhshArXiv:eq:2.18})--(\ref{zhshArXiv:eq:2.20})
we obtain the solution on isochrone
\begin{equation}\label{zhshArXiv:eq:2.23}
R^1(x,t_*)=R^1_0(b(\tau)), \quad R^2(x,t_*)=R^2_0(a(\tau)), \quad
x=X(\tau).
\end{equation}
Moving along isochrone, that is, changing the parameter $\tau$, we obtain the solution which depends on  $x$ as the fixed time moment $t=t_*$. Pay  attention that the right hand sides of differential equations, in particular, $t_a(a,b)$, $t_b(a,b)$ are easily computed with the help of (\ref{zhshArXiv:eq:2.08}), (\ref{zhshArXiv:eq:2.09}), (\ref{zhshArXiv:eq:2.11}).

We present some auxiliary notation and relations that are useful in calculations
\begin{equation}\label{zhshArXiv:eq:2.24}
\varphi(\tau|a,b)=\frac12\varphi(R^1_0(\tau),R^2_0(\tau)|r^1(b),r^2(a)).
\end{equation}
Then relation (\ref{zhshArXiv:eq:2.11}) has the form
\begin{equation}\label{zhshArXiv:eq:2.25}
t(a,b)=t_0+\int\limits_{a}^{b}\varphi(\tau|a,b)\,d\tau.
\end{equation}
The right hand sides of equation (\ref{zhshArXiv:eq:2.18}) can be easily computed
\begin{equation}\label{zhshArXiv:eq:2.26}
t_a(a,b)=-\varphi(a|a,b)+\int\limits_{a}^{b}\varphi_a(\tau|a,b)\,d\tau,\quad
t_b(a,b)=\varphi(b|a,b)+\int\limits_{a}^{b}\varphi_b(\tau|a,b)\,d\tau,
\end{equation}
where
\begin{equation}\label{zhshArXiv:eq:2.27}
\varphi_a(\tau|a,b)=\frac12\varphi^t_{r^2}(R^1_0(\tau),R^2_0(\tau)|r^1(b),r^2(a))r^2_a(a),
\end{equation}
\begin{equation}\label{zhshArXiv:eq:2.28}
\varphi_b(\tau|a,b)=\frac12\varphi^t_{r^1}(R^1_0(\tau),R^2_0(\tau)|r^1(b),r^2(a))r^1_b(b).
\end{equation}

We make several important notations. First, the right hand sides of differential equations (\ref{zhshArXiv:eq:2.18}) can be set with accuracy to an arbitrary multiplier, which essentially override the parameter $\tau$. In some cases, a good choice of the parameter $\tau$ allows to solve the Cauchy problem more effectively. Second, one \textbf{\emph{should not}} put $\tau=x$. This replacement reduce the number of equations and give more natural form of the solution: $R^1(x,t_*)=R^1_0(b(x))$, $R^2(x,t_*)=R^2_0(a(x))$. However, this option does not allow us to construct the multi-valued solutions, in particular, does not allow  to study the breaking solutions (see \cite{Zhuk_Shir_ArXiv_2014_1}).

\setcounter{equation}{0}

\section{Zonal electrophoresis}\label{zhshArXiv:sec:3}

To demonstrate the effectiveness of the proposed method we consider the  solving of the zonal electrophoresis equations with different initial data. The results are given in terms of the original equations, i.e. for values of $u^i$, and in terms of the Riemann invariants $R^i$. This section also demonstrates the possible simplification of the method in the case when the  Riemann--Green function is represented in the form
(\ref{zhshArXiv_Part_2:eq:1.01}).

We consider the system of equations describing the process of mass transfer under the action of the electric field, more precisely, the zonal electrophoresis model \cite{ElaevaMM,Elaeva_ZhVM,BabskiiZhukovYudovichRussian,ZhukovMassTransport}
\begin{equation}\label{zhshArXiv:eq:3.01}
u^1_t+
\mu^1\mu^2\left(
\frac{\mu^1 u^1}{1+s}
\right)_x=0, \quad
u^2_t+
\mu^1\mu^2\left(
\frac{\mu^2 u^2}{1+s}
\right)_x=0, \quad s=u^1+u^2,
\end{equation}
\begin{equation}\label{zhshArXiv:eq:3.02}
u^1\bigr|_{t=0}=u^1_0(x), \quad u^2\bigr|_{t=0}=u^2_0(x).
\end{equation}
Here $u^k$ are the concentrations, $\mu^k$ are the effective component mobilities, $s$ is the conductivity of the mixture,
$u^1_0(x)$, $u^2_0(x)$ are given functions (initial concentration distributions).

This system written in the Riemann invariants has the form
\begin{equation}\label{zhshArXiv:eq:3.03}
R^1_t+ \lambda^1(R^1,R^2)R^1_x=0, \quad  R^2_t+ \lambda^2(R^1,R^2)R^2_x=0,
\end{equation}
\begin{equation}\label{zhshArXiv:eq:3.04}
\lambda^1(R^1,R^2)= R^1 (R^1R^2), \quad  \lambda^2(R^1,R^2)= R^2 (R^1R^2).
\end{equation}

The correspondence with the variables is given by the expressions
\begin{equation}\label{zhshArXiv:eq:3.05}
A_0=1+u^1+u^2, \quad    B_0=\mu^1+\mu^2+u^1\mu^2+u^2\mu^1, \quad C_0=\mu^1\mu^2,
\end{equation}
\begin{equation*}
R^1=\frac{B_0-\sqrt{D_0}}{2A_0}, \quad  R^2=\frac{B_0+\sqrt{D_0}}{2A_0}, \quad D_0=B_0^2-4A_0C_0.
\end{equation*}
\begin{equation*}
\frac{1}{R^1R^2}=\frac{1+u^1+u^2}{\mu^1\mu^2}, \quad  \frac{R^1+R^2}{R^1R^2}=\frac{\mu^1+\mu^2+u^1\mu^2+u^2\mu^1}{\mu^1\mu^2},
\end{equation*}
\begin{equation}\label{zhshArXiv:eq:3.06}
u^{1}=\frac{\mu^2(R^{1}-\mu^{1})(R^{2}-\mu^{1})}{R^{1}R^{2}(\mu^{1}-\mu^{2})},
\quad
u^{2}=\frac{\mu^1(R^{1}-\mu^{2})(R^{2}-\mu^{2})}{R^{1}R^{2}(\mu^{2}-\mu^{1})}.
\end{equation}

The initial data for (\ref{zhshArXiv:eq:3.03}), (\ref{zhshArXiv:eq:3.04}) are written with the help of (\ref{zhshArXiv:eq:3.02}) as
\begin{equation}\label{zhshArXiv:eq:3.07}
R^1\bigr|_{t=0}=R^1_0(x), \quad R^2\bigr|_{t=0}=R^2_0(x).
\end{equation}

\subsection{The Riemann--Green function and implicit solution}\label{zhshArXiv:sec:3.B}

In the case (\ref{zhshArXiv:eq:3.03}), (\ref{zhshArXiv:eq:3.04}) the equations (\ref{zhshArXiv:eq:2.03}), (\ref{zhshArXiv:eq:2.04}) have the form
\begin{equation}\label{zhshArXiv:eq:3.08}
\Phi_{R^1 R^2} + \frac{R^1}{R^2(R^1- R^2)}\Phi_{R^1}-\frac{R^2}{R^1(R^1- R^2)}\Phi_{R^2}=0.
\end{equation}
The Riemann--Green function for the equation (\ref{zhshArXiv:eq:3.08}) is well known (see, \emph{i.g.} \cite{Courant,Copson,Ibragimov})
\begin{equation}\label{zhshArXiv:eq:3.09}
\Phi(R^1,R^2|r^1,r^2)=
\frac{((R^1 + R^2)(r^1 + r^2) - 2R^1R^2 - 2r^1r^2)r^1r^2}{R^1R^2(r^1 - r^2)^2}.
\end{equation}

It is obvious that $\Phi(R^1,R^2|r^1,r^2)$ can be represent in the form
\begin{equation*}
\Phi(R^1,R^2|r^1,r^2)=\sum\limits_{k=1}^{n}\mathcal{P}_k(R^1,R^2)\mathcal{Q}_k(r^1,r^2),
\end{equation*}
where
\begin{equation*}
\mathcal{P}_1(R^1,R^2)=\frac{R^1+R^2}{R^1 R^2}, \quad \mathcal{Q}_1(r^1,r^2)=\frac{(r^1 + r^2)r^1r^2}{(r^1 - r^2)^2},
\end{equation*}
\begin{equation*}
\mathcal{P}_2(R^1,R^2)=\frac{1}{R^1 R^2}, \quad \mathcal{Q}_2(r^1,r^2)=-\frac{2(r^1 r^2)^2}{(r^1 - r^2)^2},
\end{equation*}
\begin{equation*}
\mathcal{P}_3(R^1,R^2)=1, \quad \mathcal{Q}_3(r^1,r^2)=-\frac{2 r^1 r^2}{(r^1 - r^2)^2},
\end{equation*}

This representation allows to considerably simplify the calculation of the integrals in the expressions (\ref{zhshArXiv:eq:2.25}), (\ref{zhshArXiv:eq:2.26}). In fact, the integrals are calculated only from functions $\mathcal{P}_k(R^1_0(\tau),R^2_0(\tau))$ that contain only the initial data and do not contain the variables $r^1$, $r^2$. Moreover, for problem (\ref{zhshArXiv:eq:3.01}), (\ref{zhshArXiv:eq:3.02}) the functions $\mathcal{P}_1(R^1_0(\tau),R^2_0(\tau))$, $\mathcal{P}_2(R^1_0(\tau),R^2_0(\tau))$ are linear combinations of the initial data $u^1_0$, $u^2_0$ (see (\ref{zhshArXiv:eq:3.05})).

Omitting the cumbersome calculations, we present the final form of the implicit solution for (\ref{zhshArXiv:eq:3.03}),(\ref{zhshArXiv:eq:3.04})
\begin{equation}\label{zhshArXiv:eq:3.10}
t=t(a,b)=t_0+M^{t0}(b-a)+M^{tF}F+M^{tG}G, \quad t_0=0,
\end{equation}
\begin{equation}\label{zhshArXiv:eq:3.11}
x=x(a,b)=\frac{a+b}{2}+M^{x0}(b-a)+M^{xF}F+M^{xG}G,
\end{equation}
where
\begin{equation}\label{zhshArXiv:eq:3.12}
F(a,b)=
\int\limits_{a}^{b}f(\tau)\,d\tau, \quad
f(\tau)=\frac{R^1_0(\tau)+R^2_0(\tau)}{R^1_0(\tau) R^2_0(\tau)},
\end{equation}
\begin{equation*}
G(a,b)=
\int\limits_{a}^{b}g(\tau)\,d\tau, \quad
g(\tau)=\frac{1}{R^1_0(\tau) R^2_0(\tau)},
\end{equation*}
\begin{equation}\label{zhshArXiv:eq:3.13}
M^{tF}=-\frac{r^1+r^2}{(r^1-r^2)^3}, \quad
M^{tG}=\frac{2 r^1 r^2}{(r^1-r^2)^3}, \quad
M^{t0}=\frac{2}{(r^1-r^2)^3},
\end{equation}
\begin{equation}\label{zhshArXiv:eq:3.14}
M^{xF}=-\frac{2(r^1r^2)^2}{(r^1-r^2)^3}, \quad
M^{xG}=\frac{(r^1 + r^2) (r^1 r^2)^2}{(r^1-r^2)^3},
\end{equation}
\begin{equation*}
M^{x0}=\frac{3 (r^2)^2 r^1 -(r^2)^3 +3 (r^1)^2 r^2 - (r^1)^3}{2(r^1-r^2)^3},
\end{equation*}
\begin{equation}\label{zhshArXiv:eq:3.15}
R^1(x,t)=r^1(b)=R^1_0(b), \quad R^2(x,t)=r^2(a)=R^2_0(a).
\end{equation}
We recall that for solving of the problem on  the isochrones it is sufficient to know only the function $t(a,b)$ which is determined by the formulae
(\ref{zhshArXiv:eq:3.10}), (\ref{zhshArXiv:eq:3.12}), (\ref{zhshArXiv:eq:3.13}). The expressions
(\ref{zhshArXiv:eq:3.11}), (\ref{zhshArXiv:eq:3.14}) are given for completeness.

\subsection{The solution on isochrone}\label{zhshArXiv:sec:3.C}

To determine the solution $R^1(x,t_*)$, $R^2(x,t_*)$  on isochrone $t_*=t(a_*,b_*)$ we have the Cauchy problem for the variables $a$, $b$, $F$, $G$, $X$
instead of (\ref{zhshArXiv:eq:2.18})--(\ref{zhshArXiv:eq:2.20})
\begin{equation}\label{zhshArXiv:eq:3.16}
\frac{da}{d\tau}=-t_b, \quad \frac{db}{d\tau}=t_a, \quad \frac{dF}{d\tau}=F_{\tau}, \quad \frac{dG}{d\tau}=G_{\tau},
\quad \frac{dX}{d\tau}=X_\tau,
\end{equation}
\begin{equation}\label{zhshArXiv:eq:3.17}
a\bigr|_{\tau=0}=a_*, \quad b\bigr|_{\tau=0}=b_*, \quad
F\bigr|_{\tau=0}=F_*, \quad G\bigr|_{\tau=0}=G_*, \quad
X\bigr|_{\tau=0}=X_*.
\end{equation}
Here, the right hand sides of ODE's has the following form
\begin{equation}\label{zhshArXiv:eq:3.18}
t_a=\frac{d}{da} t(a,b,F(a,b),G(a,b)), \quad t_b=\frac{d}{db} t(a,b,F(a,b),G(a,b)),
\end{equation}
\begin{equation}\label{zhshArXiv:eq:3.19}
F_{\tau}(a,b)=f(b)t_a+f(a)t_b, \quad G_{\tau}(a,b)=g(b)t_a+g(a)t_b,
\end{equation}
\begin{equation}\label{zhshArXiv:eq:3.20}
X_\tau=(\lambda^2(r^1(b),r^2(a))-\lambda^1(r^1(b),r^2(a)))t_a t_b,
\end{equation}
\begin{equation}\label{zhshArXiv:eq:3.21}
  r^1=r^1(b)=R^1_0(b), \quad r^2=r^2(a)=R^2_0(a).
\end{equation}
Note, that all the right hand sides of the differential equations are calculated using the \textbf{explicit} formulae (\ref{zhshArXiv:eq:3.12})--(\ref{zhshArXiv:eq:3.15}) and the initial data (\ref{zhshArXiv:eq:3.07}). The function $t_a$, $t_b$ do not \textbf{contain} integrals. These integrals are replaced by new variables $F$ and $G$ which are determined by the solution of the Cauchy problem.

To determine the values $F_*$, $G_*$, $X_*$ we solve the problem
\begin{equation}\label{zhshArXiv:eq:3.22}
\frac{dY(b)}{db}=(\lambda^2(r^1(b),r^2(a_*)) t_b(a_*,b,F(a_*,b),G(a_*,b)),
\end{equation}
\begin{equation*}
\frac{dF(a_*,b)}{db}=f(b), \quad \frac{dG(a_*,b)}{db}=g(b),
\end{equation*}
\begin{equation}\label{zhshArXiv:eq:3.23}
Y(a_*)=a_*, \quad F(a_*,a_*)=0, \quad G(a_*,a_*)=0.
\end{equation}
Integrating the Cauchy problem (\ref{zhshArXiv:eq:3.22}), (\ref{zhshArXiv:eq:3.23}) from $b=a_*$ to $b=b_*$ we  get the values $X_*=Y(b_*)$,
$F_*=F(a_*,b_*)$, $G_*=G(a_*,b_*)$.

Strictly speaking, the Cauchy problem (\ref{zhshArXiv:eq:3.22}), (\ref{zhshArXiv:eq:3.23}) just allows to calculate the integrals $F(a_*,b_*)$, $G(a_*,b_*)$ and to determine the correspondence between the parameter $\tau$ and $x$ coordinate. If the explicit expression (\ref{zhshArXiv:eq:3.11}) for the function $x(a,b)$ is given then one can exclude the equations for $X(\tau)$ from the Cauchy problem.

In the next sections,  the numerical solution of the problem (\ref{zhshArXiv:eq:3.01}), (\ref{zhshArXiv:eq:3.02}) or problem (\ref{zhshArXiv:eq:3.03}), (\ref{zhshArXiv:eq:3.04}), (\ref{zhshArXiv:eq:3.07}) with the different initial data   are presented.

\subsubsection{The periodic initial data}\label{zhshArXiv:sec:3.C.1}

We assume that the concentration of $u^i$ as time $t=0$ is periodic in space
\begin{equation}\label{zhshArXiv:eq:3.24}
u^1_0(x)=u^1_*+\gamma^1\cos \Omega^1 x, \quad u^2_0(x)=u^2_*+\gamma^2\sin \Omega^2 x,
\end{equation}
where $u^i_*$, $\gamma^i$, $\Omega^i$ are constants.

Physically these initial concentration distributions correspond to a periodic perturbation of constant concentrations $u^1_*$, $u^2_*$ with amplitudes
$\gamma^1$, $\gamma^2$ and periods $2\pi/\Omega^1$, $2\pi/\Omega^2$.

The results of calculations are given for parameters
\begin{equation}\label{zhshArXiv:eq:3.25}
\mu^1=1.0, \quad \mu^2=3.0, \quad u^1_*=1.0, \quad u^2_*=4.0,
\end{equation}
\begin{equation*}
\gamma^1=0.3, \quad \gamma^2=0.9, \quad \Omega^1=1.0, \quad \Omega^2=2.0.
\end{equation*}

\begin{figure}[H]
\centering
\includegraphics[scale=1.0]{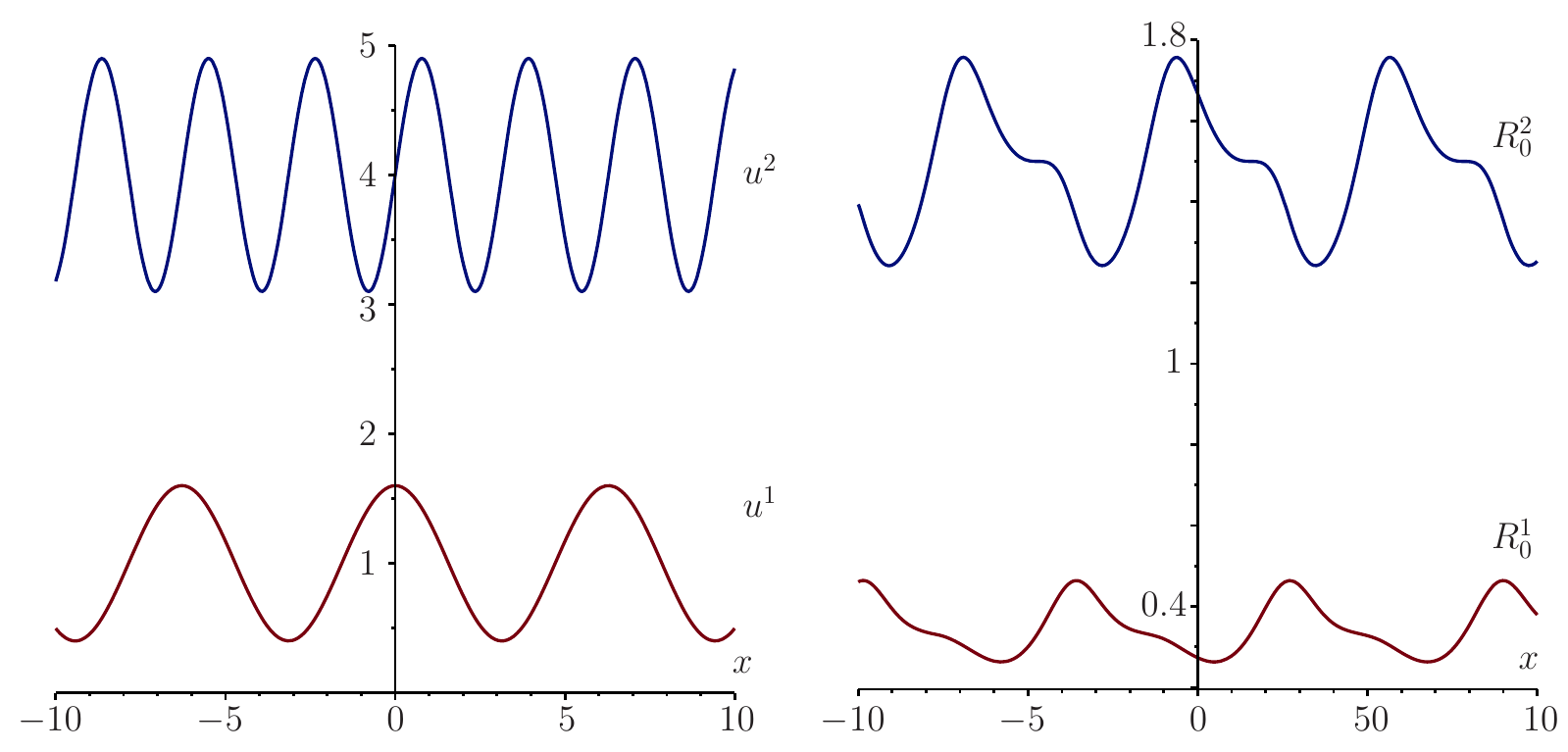}
\caption{Initial concentrations  $u^1_0(x)$, $u^2_0(x)$ and the Riemann invariants $R^1_0(x)$, $R^2_0(x)$}
\label{fig11.1.1}
\end{figure}

The initial concentration distributions and the corresponding distribution of the Riemann invariants are shown on Fig.~\ref{fig11.1.1}.

\begin{figure}[H]
\centering
\includegraphics[scale=1.0]{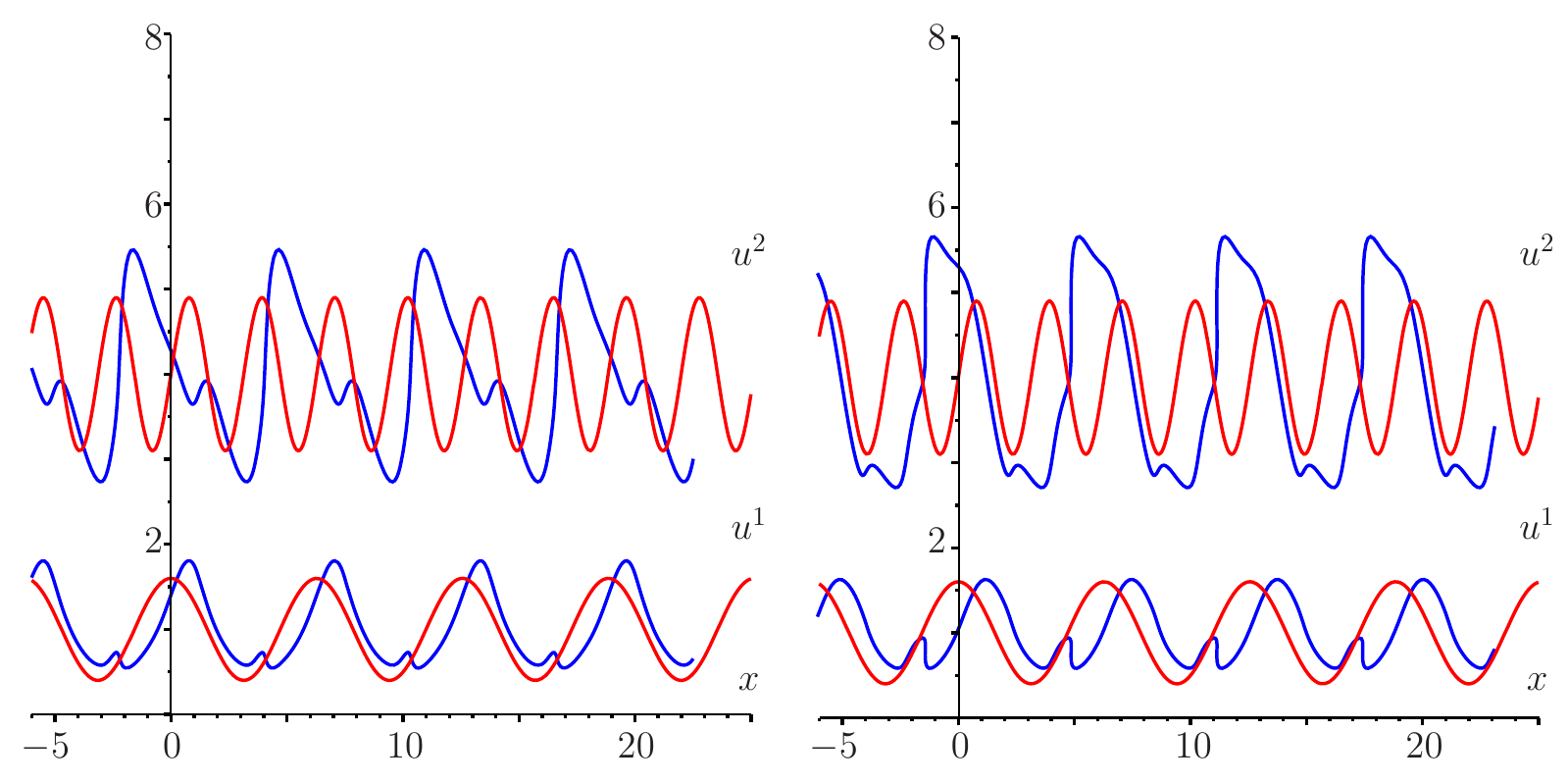}
\caption[The concentrations at $t(a_*,b_*)=1.744$ and $t(a_*,b_*)=2.944$]{
(left)  $a_*=-0.29$, $b_*=0.90$, $\delta \tau=0.05 $, $x(a_*,b_*)= 1.106$, $t(a_*,b_*)=1.744$;   
(right) $a_*=-1.09$, $b_*=0.90$, $\delta \tau=0.05 $, $x(a_*,b_*)= 1.260$, $t(a_*,b_*)=2.944$    
}
\label{fig11.1.6}
\end{figure}

\begin{figure}[H]
\centering
\includegraphics[scale=1.0]{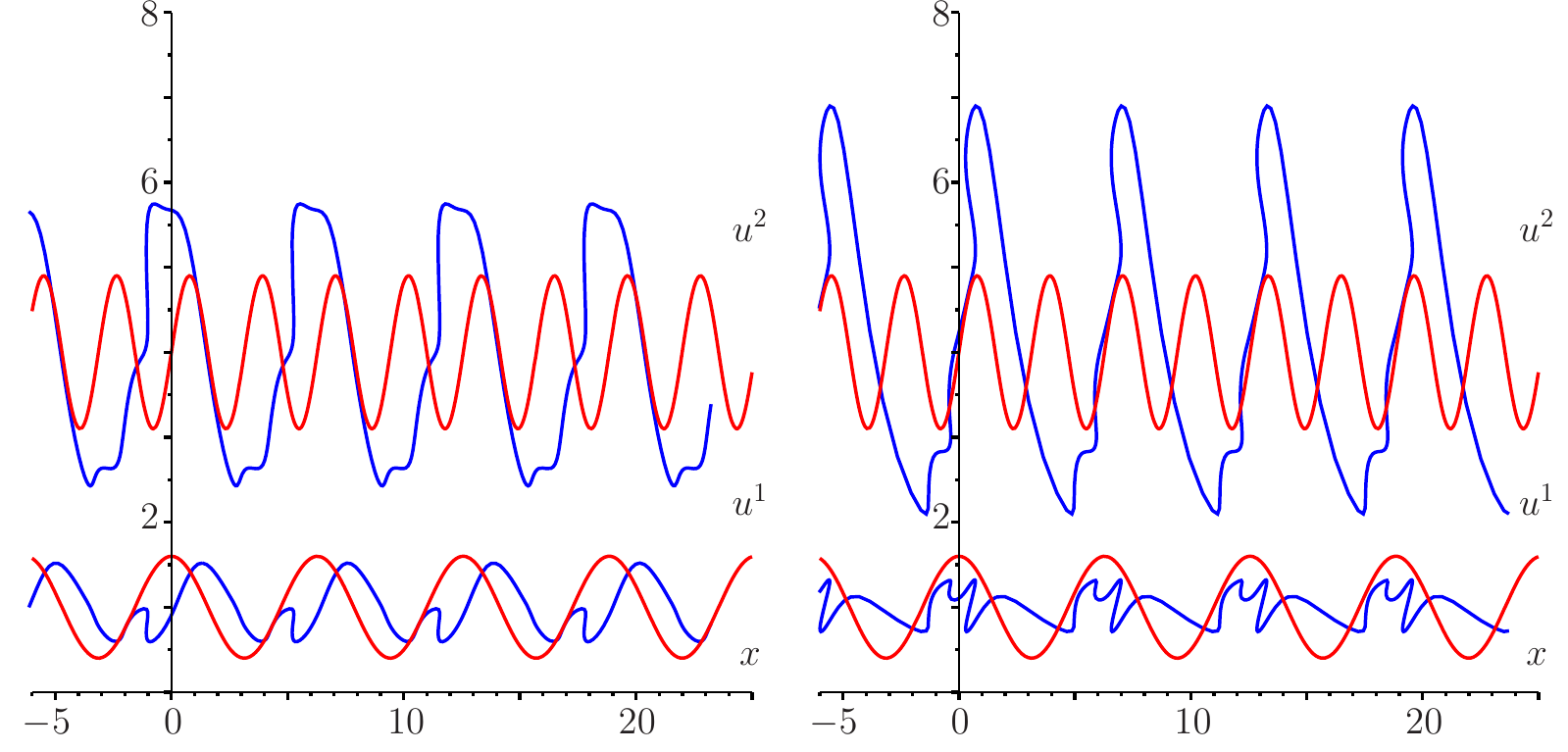}
\caption[The concentrations at $t(a_*,b_*)=3.536$ and $t(a_*,b_*)=6.174$]{
(left)  $a_*=-1.29$, $b_*=0.90$, $\delta \tau=0.05 $, $x(a_*,b_*)= 1.332$, $t(a_*,b_*)=3.536$;  
(right) $a_*=-1.89$, $b_*=0.90$, $\delta \tau=0.05 $, $x(a_*,b_*)= 1.623$, $t(a_*,b_*)=6.174$   
}
\label{fig11.1.7}
\end{figure}

\begin{figure}[H]
\centering
\includegraphics[scale=1.0]{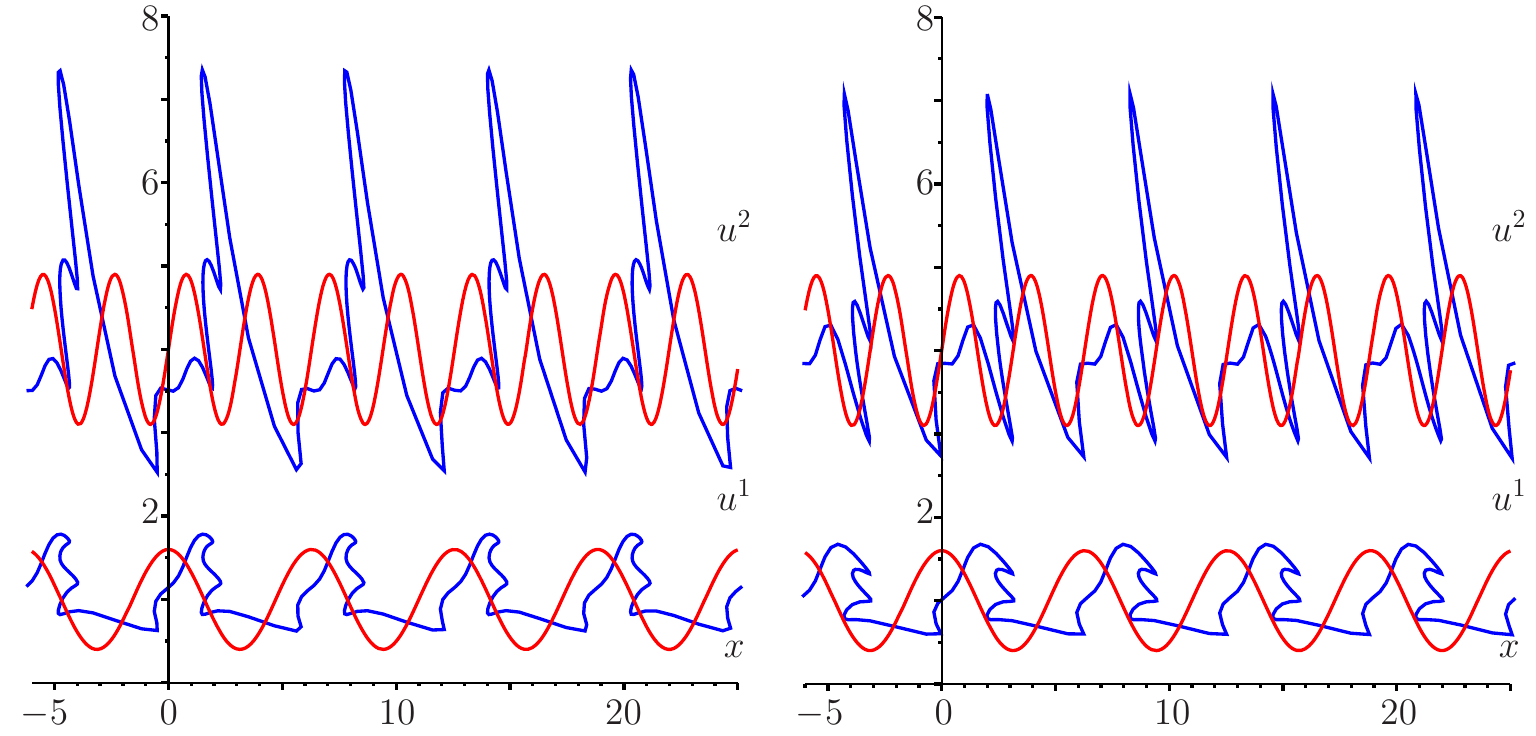}
\caption[The concentrations at $t(a_*,b_*)=8.843$ and $t(a_*,b_*)=10.137$]{
(left)  $a_*=-6.89$, $b_*=0.90$, $\delta \tau=0.05 $, $x(a_*,b_*)= 1.875$, $t(a_*,b_*)=8.843$;  
(right)  $a_*=-2.89$, $b_*=0.90$, $\delta \tau=0.05 $, $x(a_*,b_*)= 2.004$, $t(a_*,b_*)=10.137$  
}
\label{fig11.1.8}
\end{figure}

\begin{figure}[H]
\centering
\includegraphics[scale=1.0]{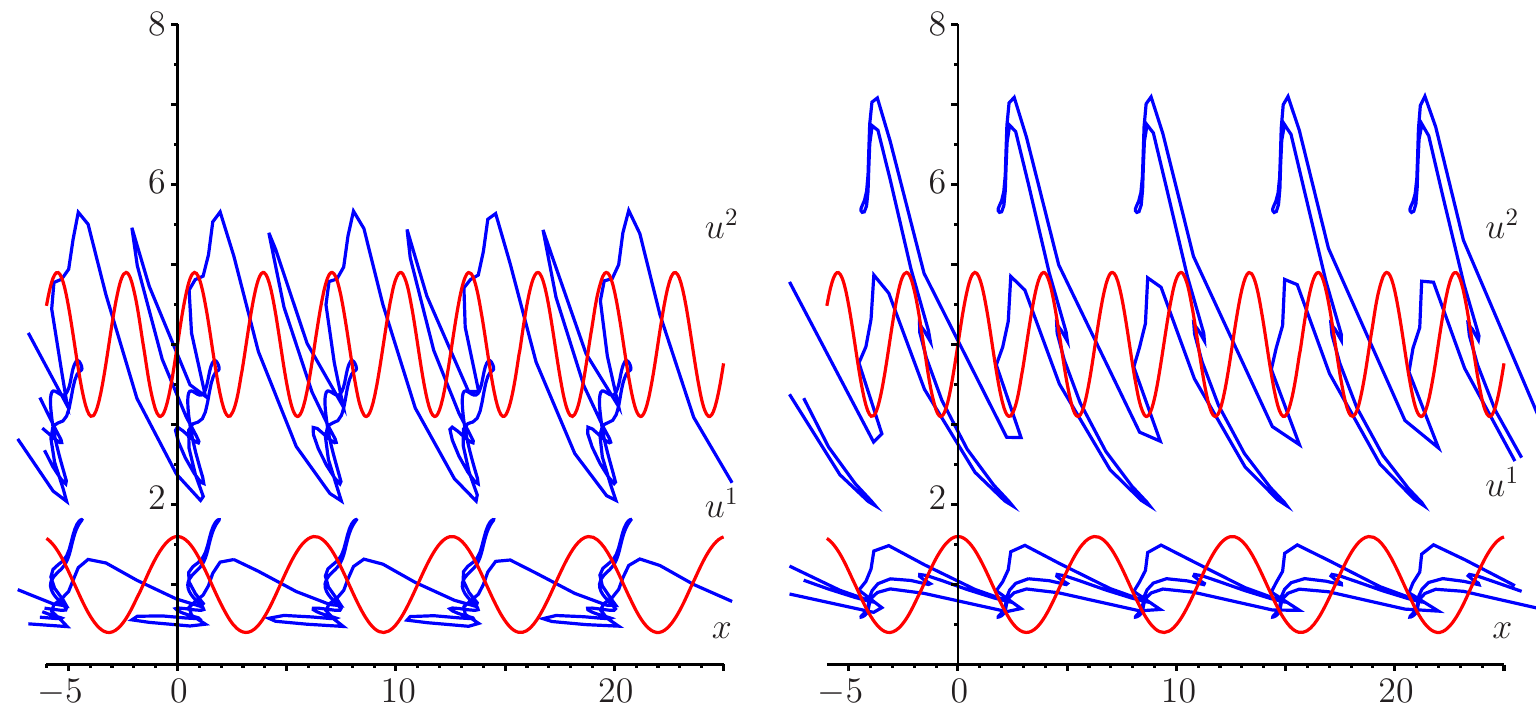}
\caption[The concentrations at $t(a_*,b_*)=14.728$ and $t(a_*,b_*)=21.120$]{
(left)    $a_*=-7.89$, $b_*=0.90$, $\delta \tau=0.05 $, $x(a_*,b_*)= 2.580$, $t(a_*,b_*)=14.728$;   
(right)   $a_*=-9.89$, $b_*=0.90$, $\delta \tau=0.05 $, $x(a_*,b_*)= 3.228$, $t(a_*,b_*)=21.120$    
}
\label{fig11.1.9}
\end{figure}

The results of calculations of the functions $u^1(x,t)$, $u^2(x,t)$ are shown on Fig.~\ref{fig11.1.6}--\ref{fig11.1.9} as $t=1.744$, $2.944$,
$3.536$, $6.174$, $8.843$, $10.137$, $14.728$, $21.120$. The red lines correspond to  the initial periodic distribution, and the blue lines correspond to the concentration distribution at the appropriate time. For all calculations we use  parameter step $\delta\tau=0.05$. We recall that the values  $a_*$, $b_*$ are correspond to isochrone $t_*=t(a_*,b_*)$ .

In a real physical situation concentrations $u^1(x,t)$, $u^2(x,t)$ are one-valued functions of the variable $x$. Thus, the real situation corresponds to time  before the moment of the breaking concentration profiles (approximately, $t\approx 3.536$, the first three figires). The remaining figures illustrate the possibilities of the method which allows to construct the multi-valued solutions.

\begin{figure}[H]
\centering
\includegraphics[scale=1.0]{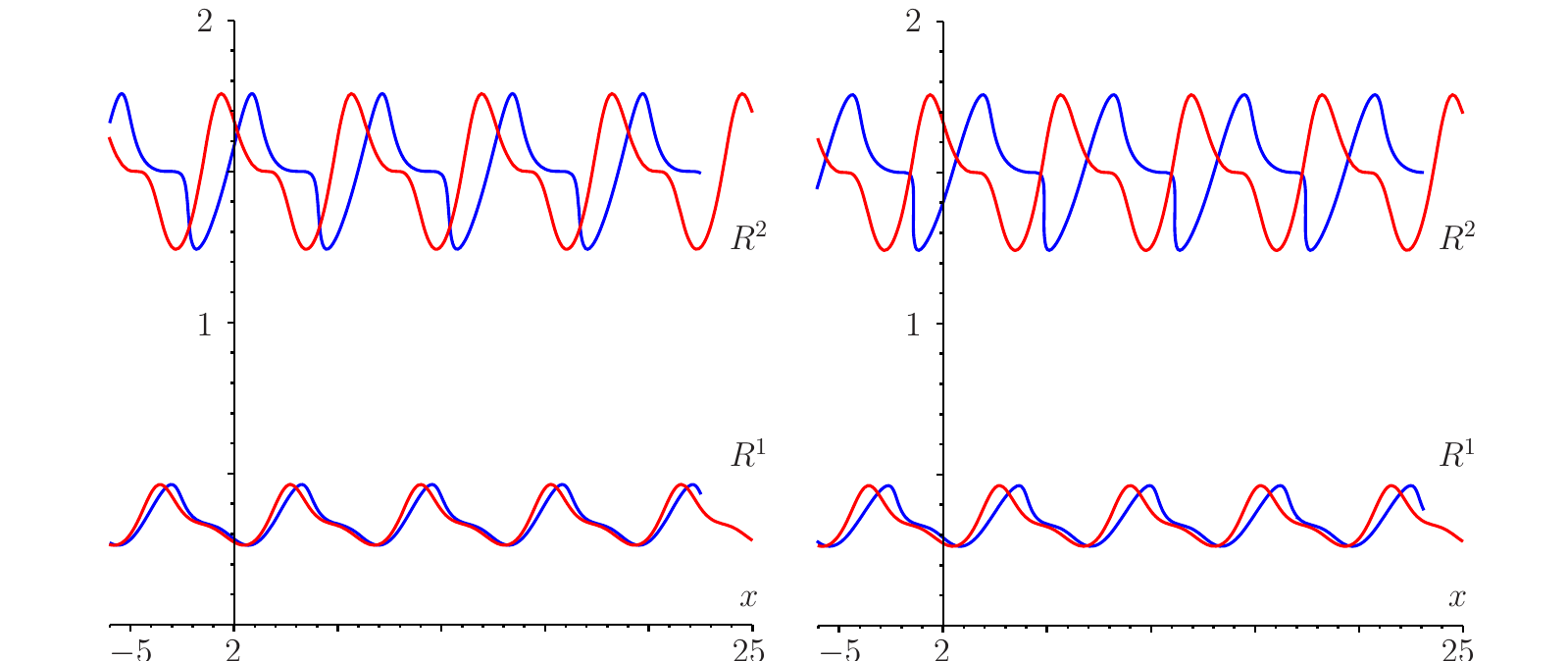}
\caption[The Riemann invariants at $t(a_*,b_*)=1.744$, $t(a_*,b_*)=2.944$]{
(left)  $a_*=-0.29$, $b_*=0.90$, $\delta \tau=0.05 $, $x(a_*,b_*)= 1.106$, $t(a_*,b_*)=1.744$;   
(right) $a_*=-1.09$, $b_*=0.90$, $\delta \tau=0.05 $, $x(a_*,b_*)= 1.260$, $t(a_*,b_*)=2.944$    
}
\label{fig11.1.12}
\end{figure}

\begin{figure}[H]
\centering
\includegraphics[scale=1.0]{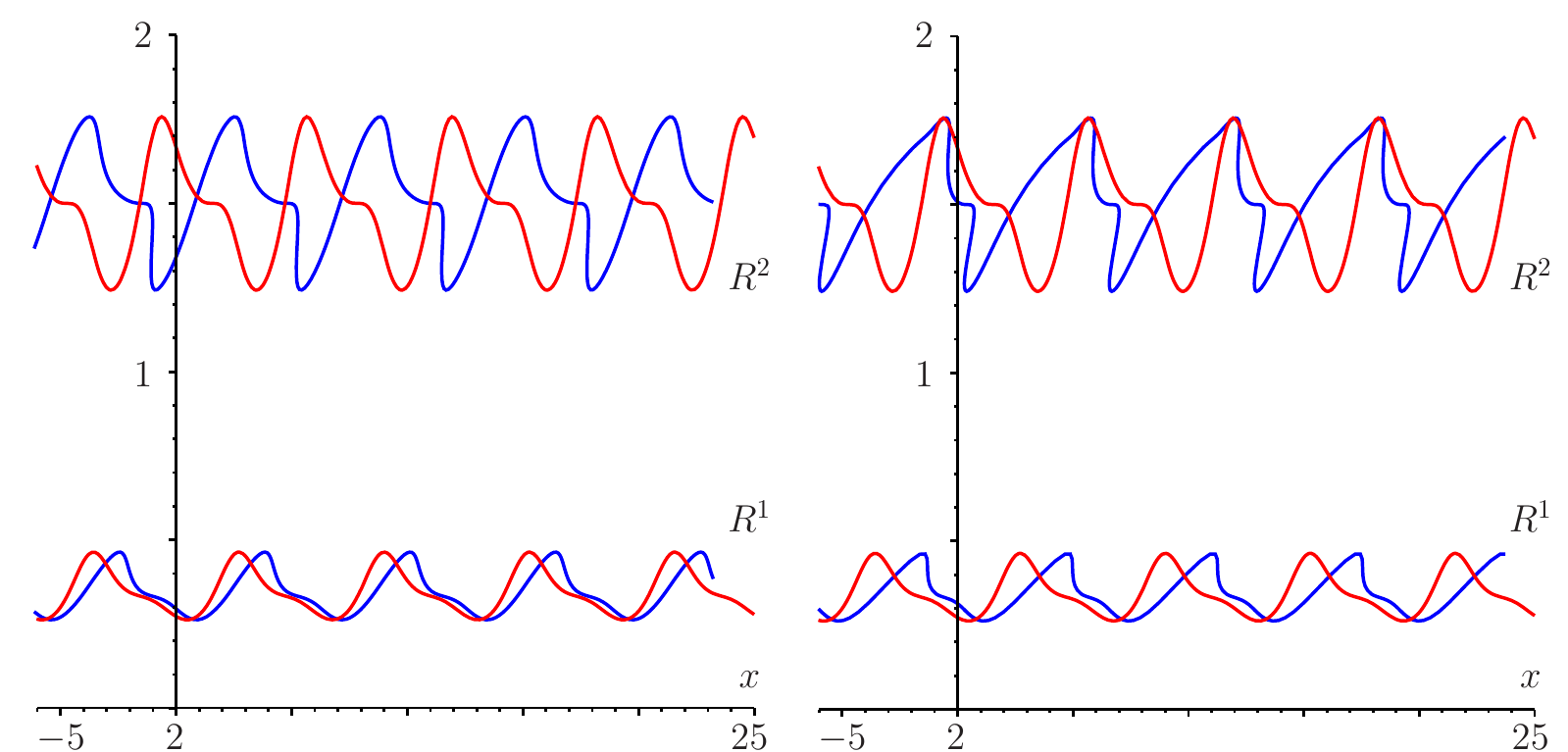}
\caption[The Riemann invariants at $t(a_*,b_*)=3.536$, $t(a_*,b_*)=6.174$]{
(left)  $a_*=-1.29$, $b_*=0.90$, $\delta \tau=0.05 $, $x(a_*,b_*)= 1.332$, $t(a_*,b_*)=3.536$;  
(right) $a_*=-1.89$, $b_*=0.90$, $\delta \tau=0.05 $, $x(a_*,b_*)= 1.623$, $t(a_*,b_*)=6.174$   
}
\label{fig11.1.13}
\end{figure}

\begin{figure}[H]
\centering
\includegraphics[scale=1.0]{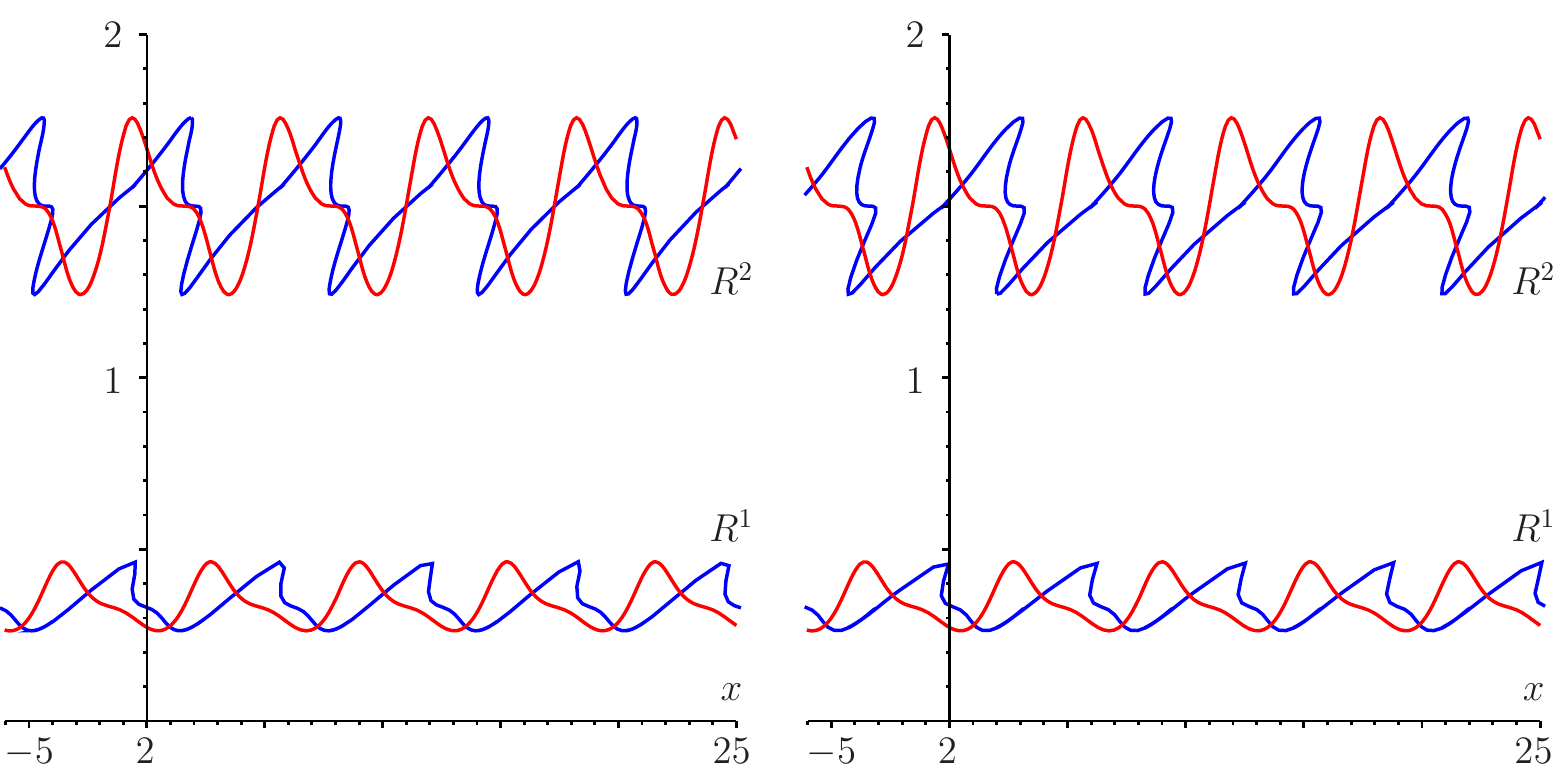}
\caption[The Riemann invariants at $t(a_*,b_*)=8.843$, $t(a_*,b_*)=10.137$]{
(left)   $a_*=-6.89$, $b_*=0.90$, $\delta \tau=0.05 $, $x(a_*,b_*)= 1.875$, $t(a_*,b_*)=8.843$;  
(right)  $a_*=-2.89$, $b_*=0.90$, $\delta \tau=0.05 $, $x(a_*,b_*)= 2.004$, $t(a_*,b_*)=10.137$  
}
\label{fig11.1.14}
\end{figure}

\begin{figure}[H]
\centering
\includegraphics[scale=1.0]{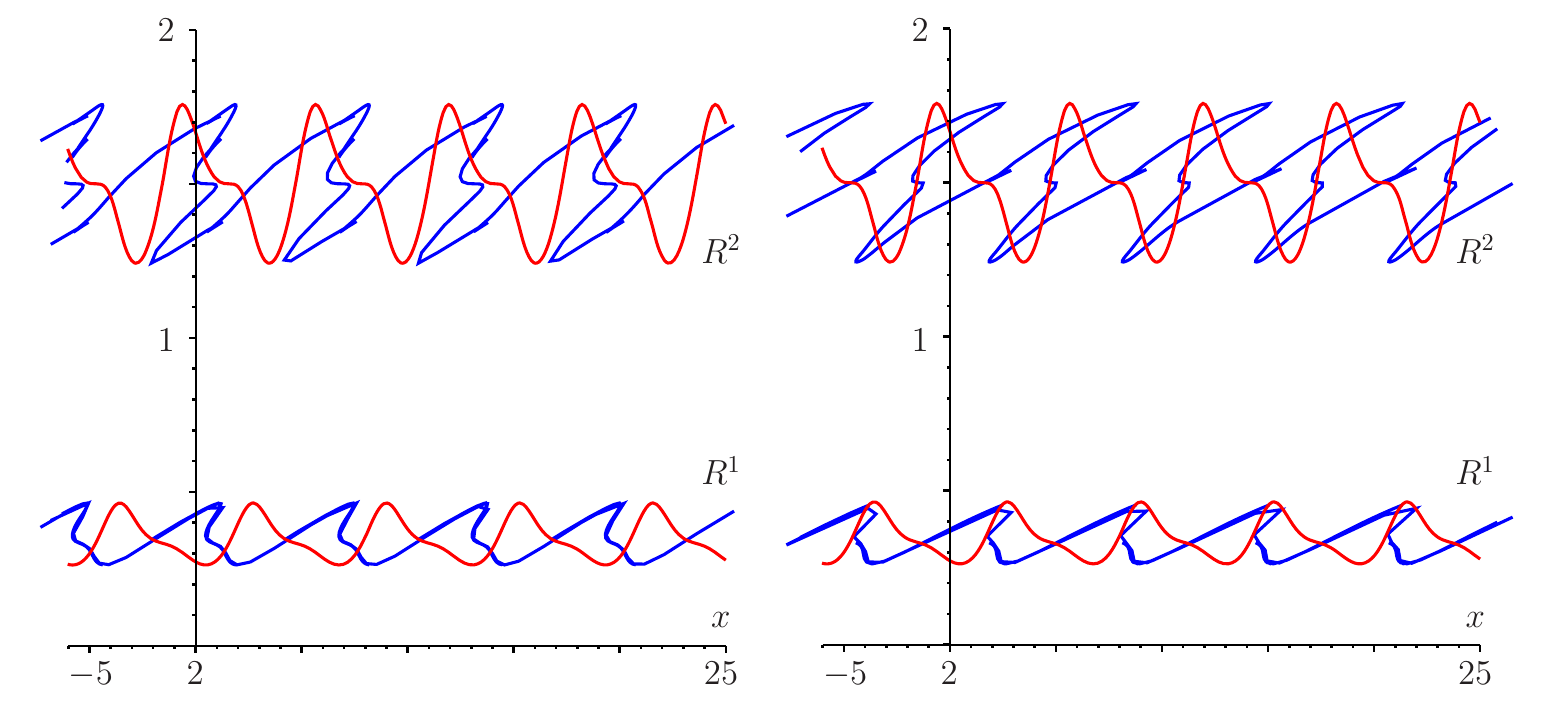}
\caption[The Riemann invariants at $t(a_*,b_*)=14.728$, $t(a_*,b_*)=21.120$]{
(left)    $a_*=-7.89$, $b_*=0.90$, $\delta \tau=0.05 $, $x(a_*,b_*)= 2.580$, $t(a_*,b_*)=14.728$;   
(right)   $a_*=-9.89$, $b_*=0.90$, $\delta \tau=0.05 $, $x(a_*,b_*)= 3.228$, $t(a_*,b_*)=21.120$    
}
\label{fig11.1.15}
\end{figure}

The evolution of the Riemann invariants $R^1(x,t)$, $R^2(x,t)$ corresponding to Fig.~\ref{fig11.1.6}--\ref{fig11.1.9} is shown on Fig.~\ref{fig11.1.12}--\ref{fig11.1.15}. Note that changing of the Riemann invariants $R^1$, $R^2$ over time is more `regular' than changing of concentrations $u^1$, $u^2$.

\subsubsection{Wave packet}\label{zhshArXiv:sec:3.C.2}

We assume that the concentrations $u^i$ as time $t=0$ are
\begin{equation}\label{zhshArXiv:eq:3.26}
u^1_0=u^1_*(1+\gamma^1e^{-\beta^1 x^2}\cos\Omega^1 x),
\end{equation}
\begin{equation*}
u^2_0=u^2_*(1+\gamma^2e^{-\beta^2 x^2}\sin\Omega^2 x),
\end{equation*}
where $u^i_*$,  $\gamma^i$, $\Omega^i$, $\beta^i$ are the constants.

Physically this initial concentration distribution corresponds to the  wave packet perturbation.

The results of calculations are given for parameters
\begin{equation}\label{zhshArXiv:eq:3.27}
\mu^1=1.0, \quad \mu^2=3.0, \quad u^1_*=1.0, \quad u^2_*=4.0,
\quad \beta^1=\beta^2=0.1,
\end{equation}
\begin{equation*}
\gamma^1=0.1, \quad \gamma^2=0.3, \quad \Omega^1=5.0, \quad \Omega^2=5.0.
\end{equation*}

\begin{figure}[H]
\centering
\includegraphics[scale=1.0]{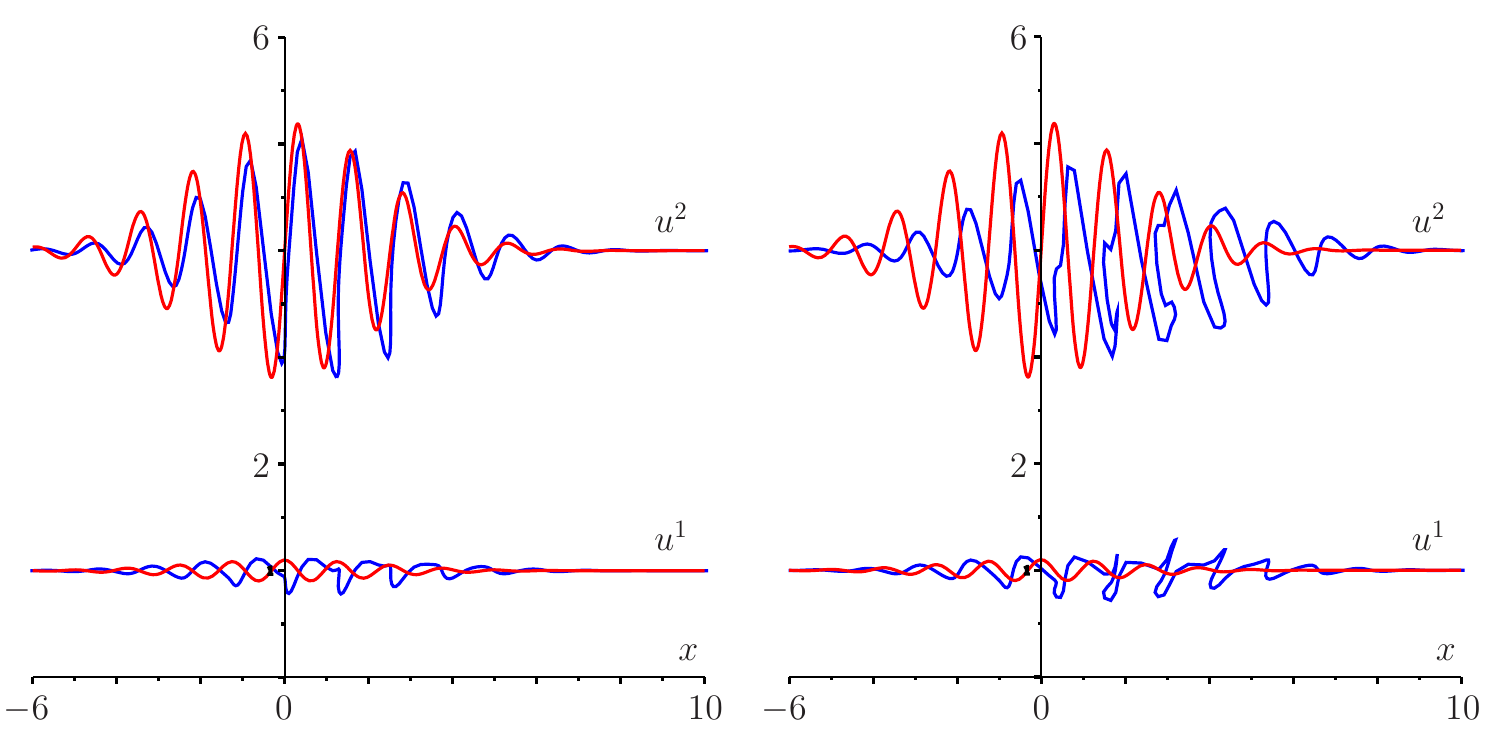}
\caption[Wave packet]{
(left)    $a_*=-0.29$, $b_*=0.90$, $\delta \tau=0.05 $, $x(a_*,b_*)= 1.238$, $t(a_*,b_*)=1.655$;
(right)   $a_*=-1.29$, $b_*=0.90$, $\delta \tau=0.05 $, $x(a_*,b_*)= 1.673$, $t(a_*,b_*)=3.737$
}
\label{fig11.2.1}
\end{figure}

\begin{figure}[H]
\centering
\includegraphics[scale=1.0]{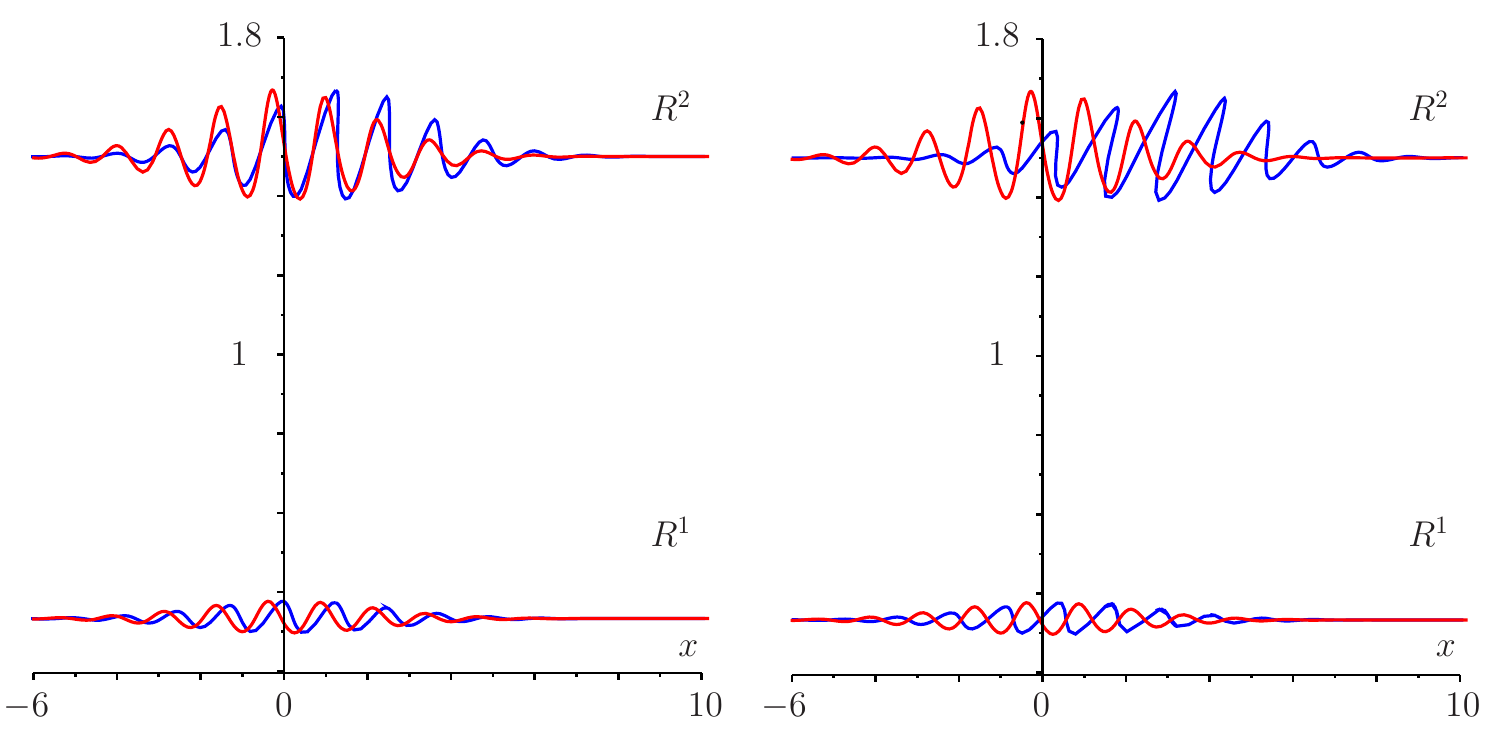}
\caption[Wave packet]{
(left)    $a_*=-0.29$, $b_*=0.90$, $\delta \tau=0.05 $, $x(a_*,b_*)= 1.238$, $t(a_*,b_*)=1.655$;
(right)   $a_*=-1.29$, $b_*=0.90$, $\delta \tau=0.05 $, $x(a_*,b_*)= 1.673$, $t(a_*,b_*)=3.737$
}
\label{fig11.2.2}
\end{figure}

We restrict only an illustration of the calculations for time $t=1.655$, $3.737$. On Fig.~\ref{fig11.2.1}, \ref{fig11.2.2} the distribution of the concentrations and the Riemann invariants, respectively, are shown.  At $t \approx 1.655$ the braking concentration profiles (and, of course, the Riemann invariants) are well visible.

\subsubsection{The Gaussian distribution  of the initial  concentrations}\label{zhshArXiv:sec:3.C.3}

The initial Gaussian distribution of the concentrations is most often realized in zonal electrophoresis experiments.

We take the following initial data
\begin{equation}\label{zhshArXiv:eq:3.28}
u^1_0=u^1_*+ \gamma^1e^{-\beta^1 (x-x^1)^2}, \quad u^2_0=u^2_*+ \gamma^2e^{-\beta^2 (x-x^2)^2},
\end{equation}
where $u^i_*$, $\gamma^i$, $x^i$, $\beta^i$ are the constants.

The results of calculations are given for parameters
\begin{equation}\label{zhshArXiv:eq:3.29}
\mu^1=1.0, \quad \mu^2=3.0, \quad u^1_*=1.0, \quad u^2_*=4.0,
\quad \beta^1=\beta^2=0.3,
\end{equation}
\begin{equation*}
\gamma^1=2.0, \quad \gamma^2=3.0, \quad x^1=5.0, \quad x^2=0.0.
\end{equation*}

\begin{figure}[H]
\centering
\includegraphics[scale=1.0]{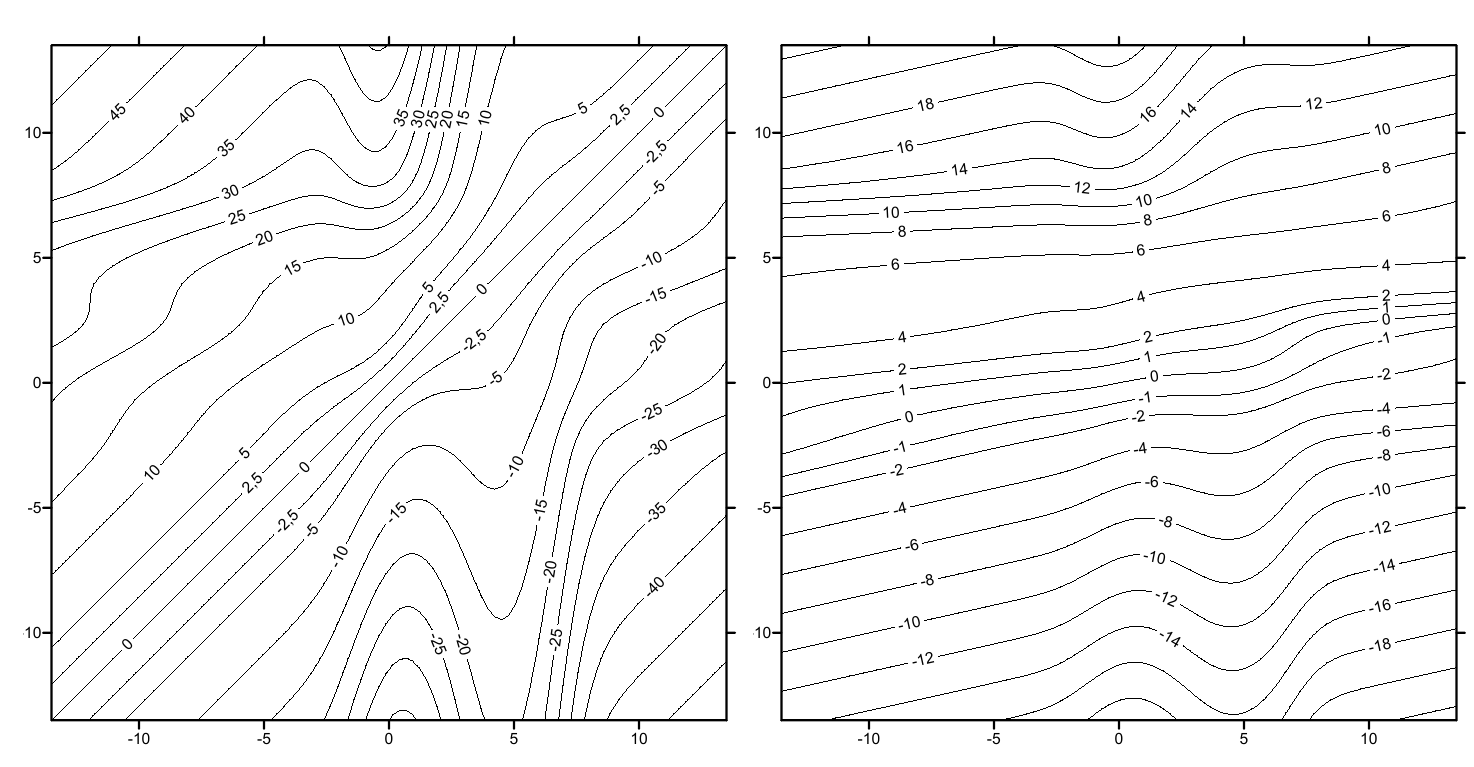}
\caption[Line levels of $t(a,b)$ and $x(a,b)$]{The isolines of the function $t(a,b)$ (left) and $x(a,b)$ (right)}
\label{fig11.3.1}
\end{figure}

On Fig.~\ref{fig11.3.1} the isolines of the functions $t(a,b)$, $x(a,b)$ on the plane $(a,b)$ are shown.

\begin{figure}[H]
\centering
\includegraphics[scale=1.0]{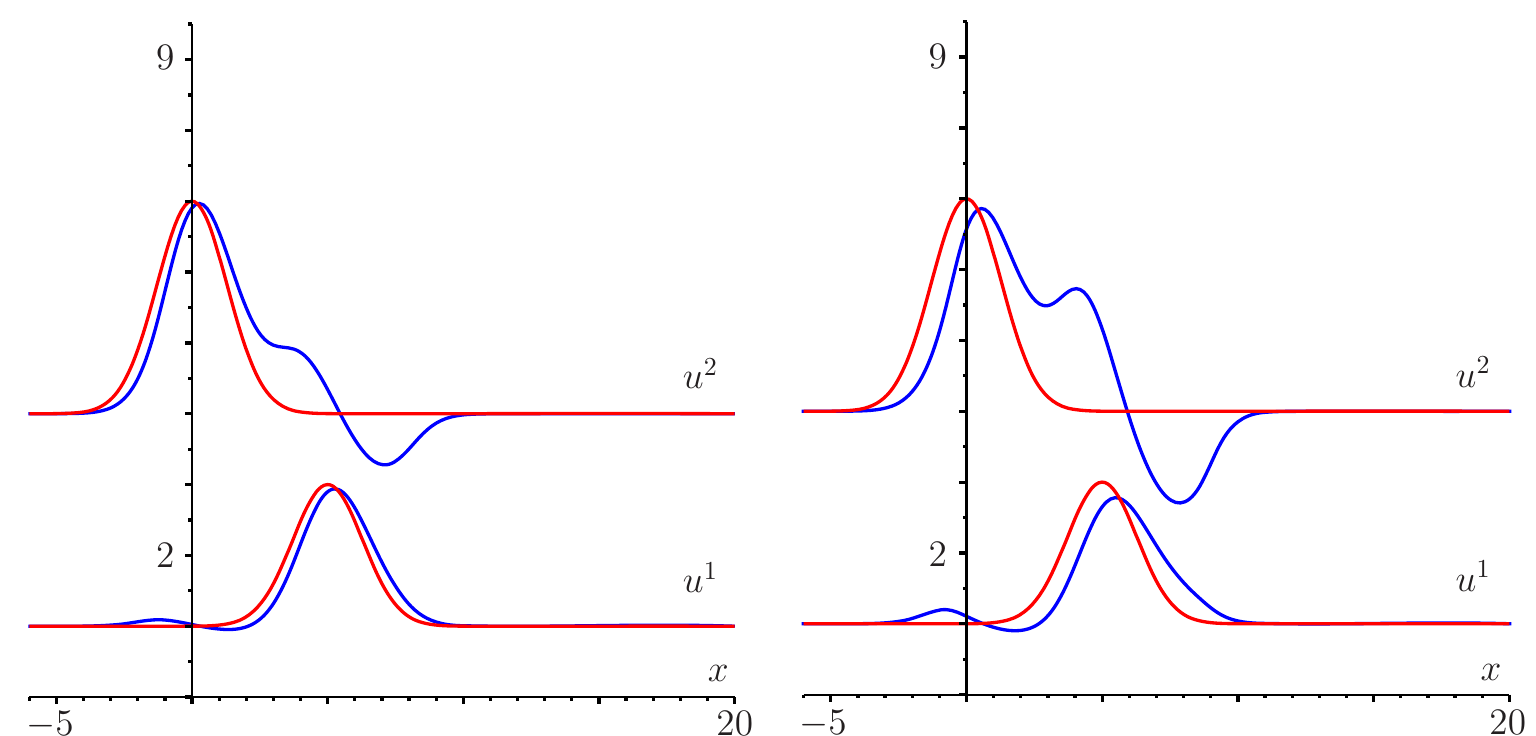}
\caption[$u^k(x,t)$ at $t(a_*,b_*)=1.146$ и $t(a_*,b_*)=2.338$]{
(left)  $a_*=0.49$, $b_*=0.90$, $\delta \tau=0.05 $, $x(a_*,b_*)= 1.009$, $t(a_*,b_*)=1.146$;    
(right) $a_*=0.09$, $b_*=0.90$, $\delta \tau=0.05 $, $x(a_*,b_*)= 1.122$, $t(a_*,b_*)=2.338$     
}
\label{fig11.3.3}
\end{figure}

\begin{figure}[H]
\centering
\includegraphics[scale=1.0]{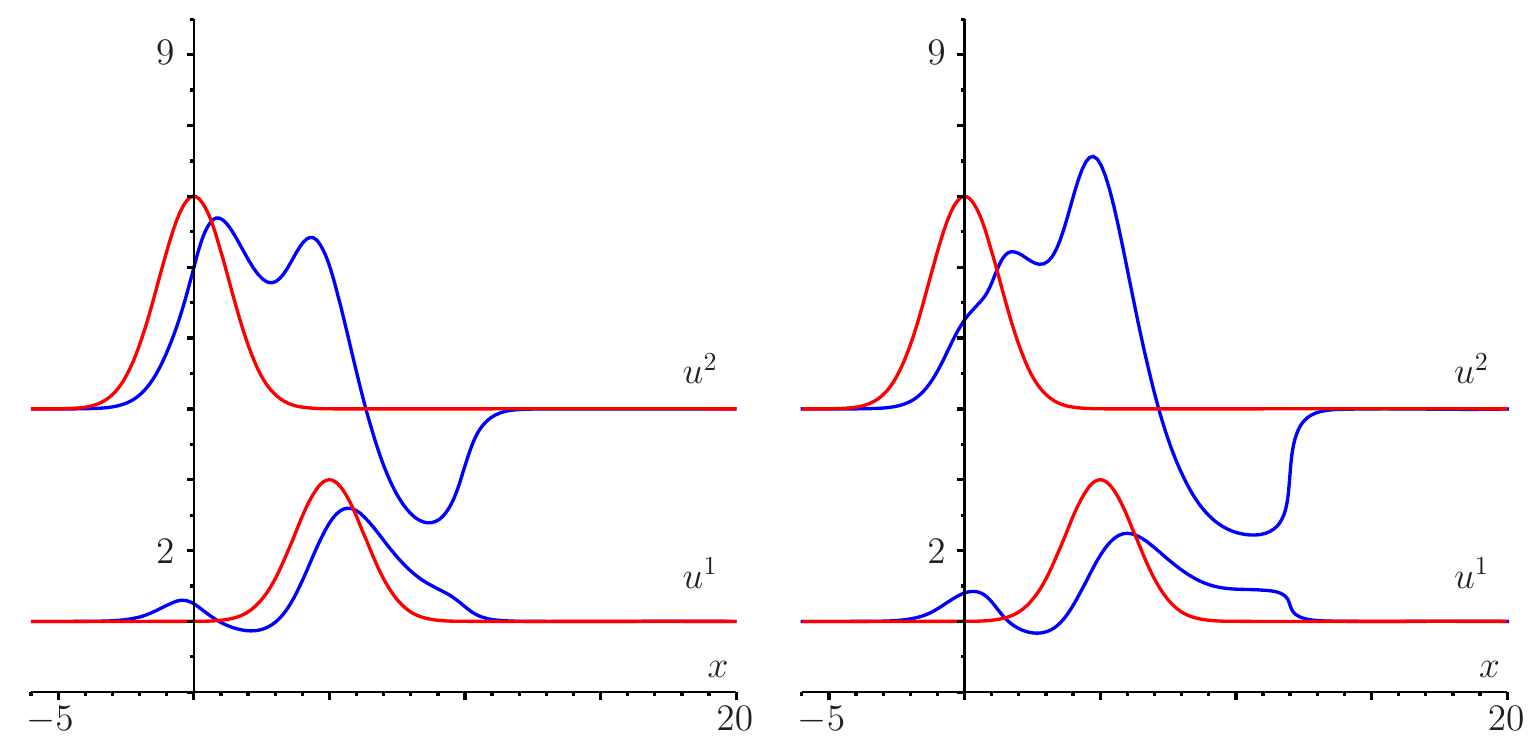}
\caption[$u^k(x,t)$ at $t(a_*,b_*)=3.438$ и $t(a_*,b_*)=5.614$]{
(left)  $a_*=-0.29$, $b_*=0.90$, $\delta \tau=0.05 $, $x(a_*,b_*)= 1.226$, $t(a_*,b_*)=3.438$;  
(right) $a_*=-1.29$, $b_*=0.90$, $\delta \tau=0.05 $, $x(a_*,b_*)= 1.435$, $t(a_*,b_*)=5.614$   
}
\label{fig11.3.4}
\end{figure}

\begin{figure}[H]
\centering
\includegraphics[scale=1.0]{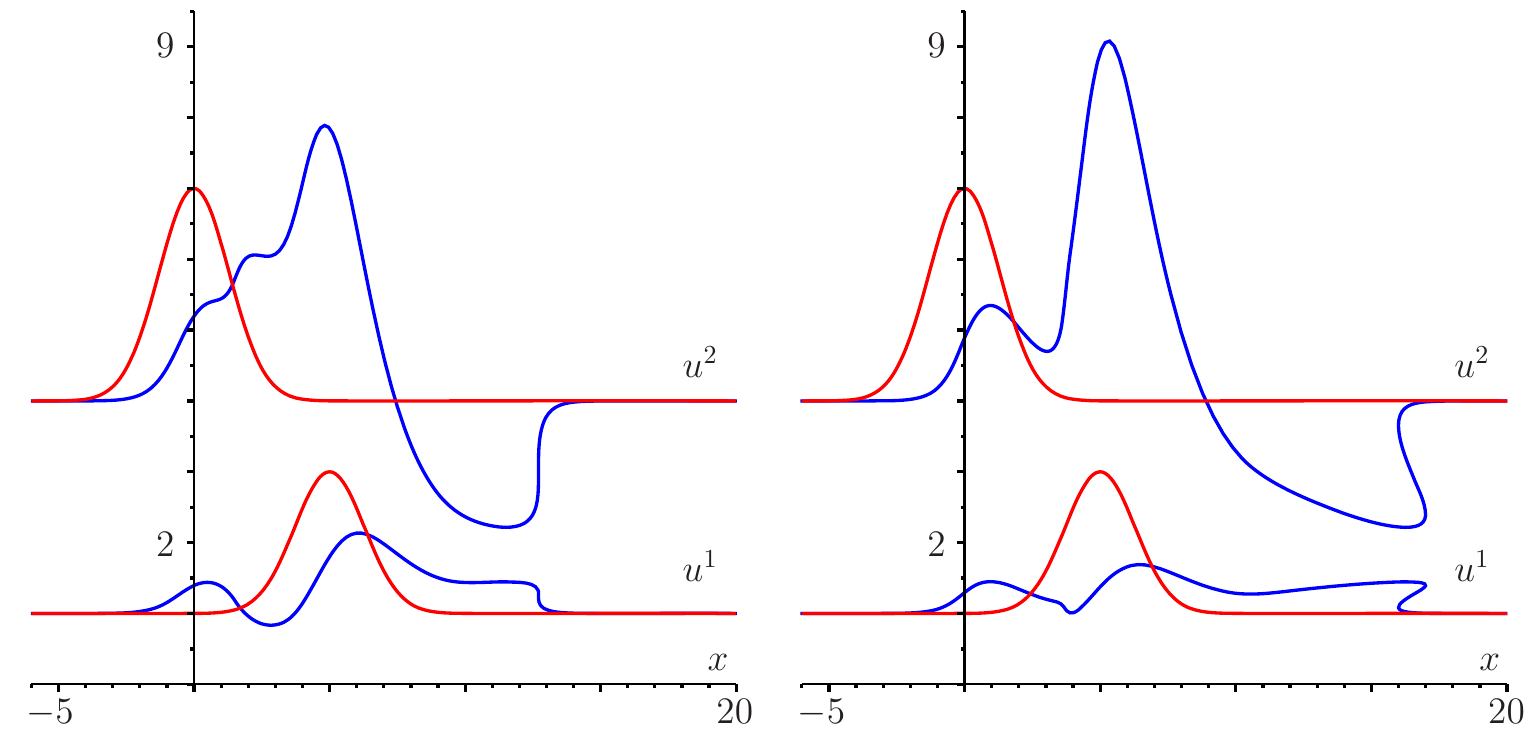}
\caption[$u^k(x,t)$ at $t(a_*,b_*)=6.358$ и $t(a_*,b_*)=10.192$]{
(left)   $a_*=-1.89$, $b_*=0.90$, $\delta \tau=0.05 $, $x(a_*,b_*)= 1.509$, $t(a_*,b_*)=6.358$;  
(right)  $a_*=-4.89$, $b_*=0.90$, $\delta \tau=0.05 $, $x(a_*,b_*)= 1.921$, $t(a_*,b_*)=10.192$  
}
\label{fig11.3.5}
\end{figure}

\begin{figure}[H]
\centering
\includegraphics[scale=1.0]{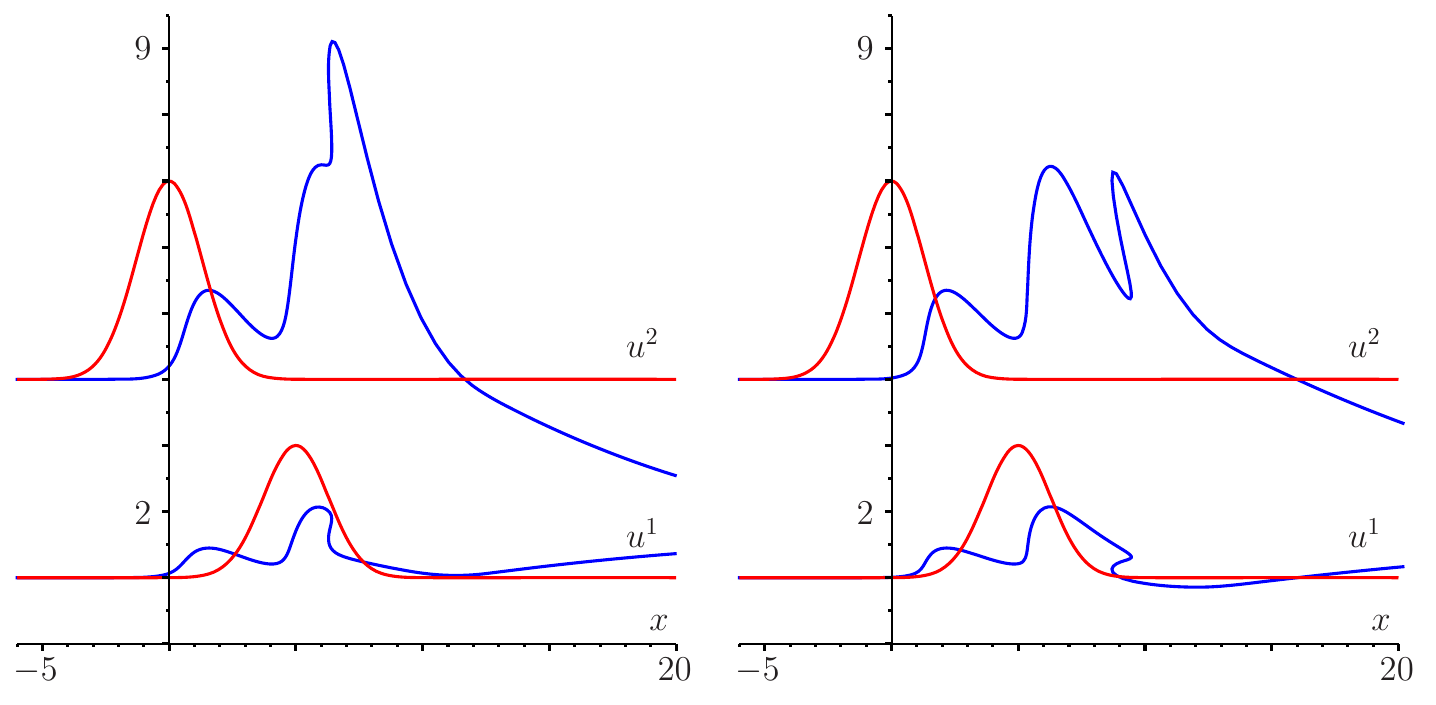}
\caption[$u^k(x,t)$ at $t(a_*,b_*)=16.350$ и $t(a_*,b_*)=22.513$]{
(left)    $a_*=-8.89$, $b_*=0.90$, $\delta \tau=0.05 $, $x(a_*,b_*)= 2.591$, $t(a_*,b_*)=16.350$;    
(right)   $a_*=-12.89$, $b_*=0.90$, $\delta \tau=0.05 $, $x(a_*,b_*)= 3.262$, $t(a_*,b_*)=22.513$    
}
\label{fig11.3.6}
\end{figure}

On Fig.~\ref{fig11.3.3}--\ref{fig11.3.6} for time moments $t=1,146$, $2.338$,$3.438$, $5.614$, $6.358$, $10.192$, $16.350$, $22.513$
the results of calculations of the functions $u^k(x,t)$ are shown.
The red lines correspond to the initial distribution, and the blue lines correspond to the concentration distribution in the appropriate time.
We use  parameter step $\delta\tau=0.05$ for all calculations. We recall that the value  $a_*$, $b_*$ are correspond to isochrone $t_*=t(a_*,b_*)$ .

We again recall that in a real physical situation the concentrations $u^1(x,t)$, $u^2(x,t)$ are the one-valued  functions of the spatial variable $x$. The real situation corresponds to  time before breaking concentration profiles (approximately, $t\approx 6.358$, the first five figures). The remaining figures illustrate the possibilities of the method which allows to construct the multi-valued solutions.

\begin{figure}[H]
\centering
\includegraphics[scale=1.0]{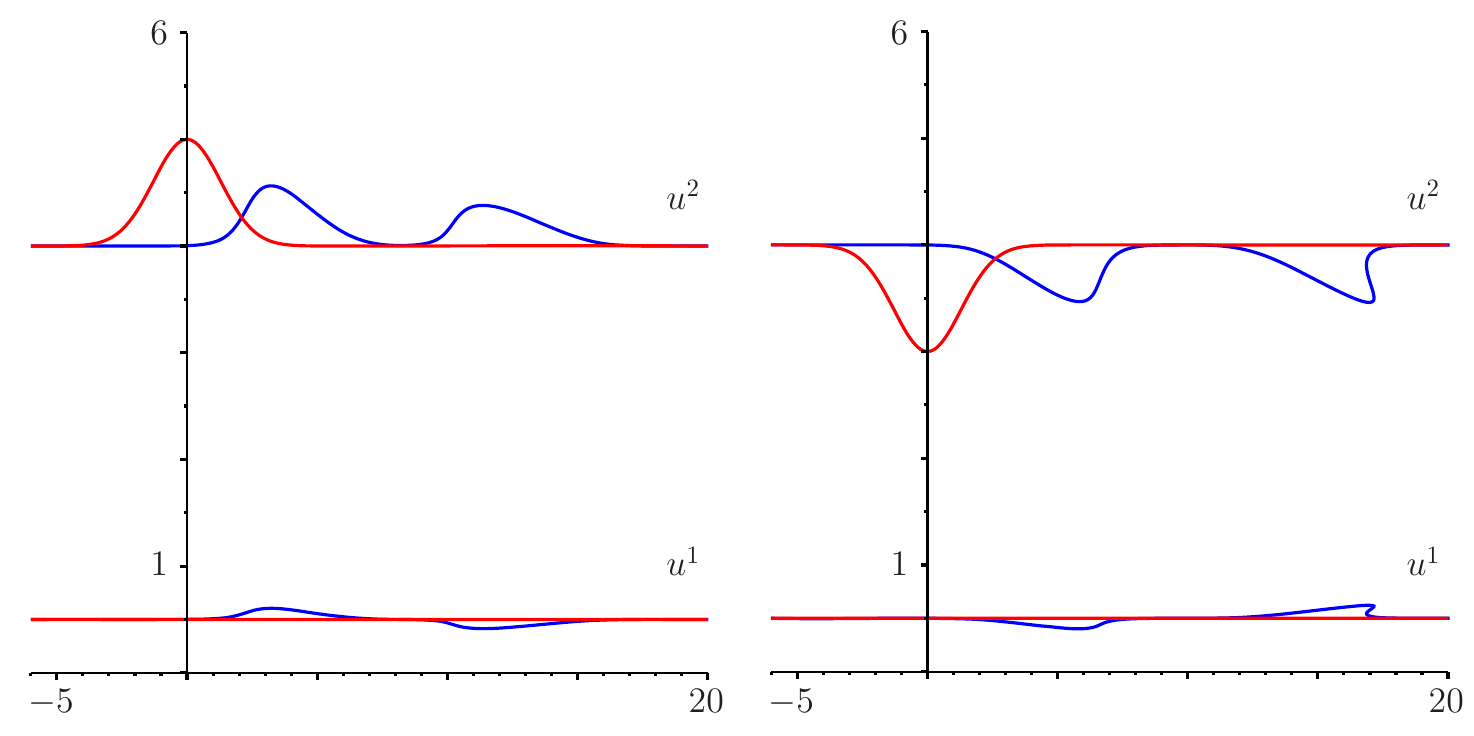}
\caption[$u^k(x,t)$]{
(left) $\gamma=+1$,    $a_*=-8.89$, $b_*=0.90$, $\delta \tau=0.05 $, $x(a_*,b_*)= 4.268$, $t(a_*,b_*)=18.091$;    
(right) $ \gamma=-1$,  $a_*=-8.89$, $b_*=0.90$, $\delta \tau=0.05 $, $x(a_*,b_*)= 6.467$, $t(a_*,b_*)=19.942$     
}
\label{fig11.3.8}
\end{figure}

The Fig.~\ref{fig11.3.8} shows the results of calculations for the case when the initial data have the form
\begin{equation}\label{zhshArXiv:eq:3.30}
u^1_0=u^1_*, \quad u^2_0=u^2_*+ \gamma e^{-\beta x^2},
\end{equation}
\begin{equation*}
\mu^1=1.0, \quad \mu^2=3.0, \quad u^1_*=0.5, \quad u^2_*=4.0,
\quad \beta=0.3, \quad \gamma=\pm 1.0.
\end{equation*}

The results illustrate that the presence of the initial perturbation of any only one concentration (in this case $u^2$) leads to the perturbation of the  other concentrations (in this case $u^1$). Strictly speaking, the result is quite obvious, as it is easy to check that the original equation (\ref{zhshArXiv:eq:3.01}) does not admit the solutions  $u^1=\const$, $u^2 \ne \const$.

It is interesting to note that the initial Gaussian's perturbation single peak, in the process of evolution, splits into two peaks. For initial data (\ref{zhshArXiv:eq:3.30}), if $\gamma=+1$, the distribution of perturbations occurs quite regularly, whereas in the case $\gamma=-1$, over time there are the breaking concentration profiles, and the occurrence of shock waves.

\setcounter{equation}{0}

\section{Conclusions}

Implicit solution (\ref{zhshArXiv:eq:3.10}), (\ref{zhshArXiv:eq:3.15}) of the original problem (\ref{zhshArXiv:eq:2.01}), (\ref{zhshArXiv:eq:2.02}), of course, is very important in studying of the original Cauchy problem properties. However, from our point of view, the implicit solution is not less complex than the original problem. For practical applications we must get, in one way or another, the explicit relations (\ref{zhshArXiv:eq:2.12}). In the general case we need to use a numerical methods, for example, Newton's method, for solving of the transcendental equations. This, in turn, requires a good initial approximations, or the using of the movement parameter method. It is especially difficult to numerically solve the system (\ref{zhshArXiv:eq:2.10}) in the case when original problem has multi-valued solition. In other words, using of the numerical methods of solving the systems of transcendental equations is not
much easier than the application of the direct methods for solving of the original problem, for example, using finite difference or finite volume method.
Instead solving of the transcendental equations system we propose to solve the Cauchy problem for ODE's. Even if it is need to solve ODE's numerically, for example, by the methods of Runge--Kutta method, the numerical algorithm is realized simpler than the algorithm for solving system of nonlinear transcendental equations.


\begin{acknowledgments}

The authors are grateful to N. M. Zhukova for proofreading the manuscript.
Funding statement. This research is partially supported by the Base Part of the Project no. 213.01-11/2014-1,
Ministry of Education and Science of the Russian Federation, Southern Federal University.

\end{acknowledgments}

\setlength{\bibsep}{4.0pt}

\end{document}